\documentclass[a4paper,11pt]{article}

\usepackage{amsmath,amssymb,amssymb,amsfonts,mathtools}
\usepackage{nicefrac}       
\usepackage[normalem]{ulem}

\usepackage{wasysym}

\usepackage{verbatim}
\usepackage{marginnote}

\usepackage[colorlinks,bookmarks,pagebackref,pdfusetitle,urlcolor=blue,citecolor=blue,linkcolor=blue,bookmarksnumbered,plainpages=false]{hyperref}
\usepackage{cleveref}

\usepackage{esint} 

\usepackage{lipsum}
\usepackage{amsfonts}

\usepackage{sectsty}

\usepackage[dvipsnames]{xcolor}
\usepackage{graphicx,psfrag,color}
\usepackage{tikz,pgfplots,psfrag}
\pgfplotsset{compat=1.18} 

\usepackage{amsopn}

\usepackage{enumerate}
\usepackage{float}
\usepackage[font=small,skip=0pt]{caption}

\usepackage{amssymb,amsmath,latexsym,amsfonts,amsthm}
\usepackage{mathtools}
\usepackage{braket}
\usepackage{float}
\usepackage{subcaption}
\usepackage{ulem}
\usepackage{cancel}
\usepackage{braket}
\usepackage{bm}
\usepackage{cases}
\usepackage{empheq}

\usepackage{graphicx,color}
\usepackage{epstopdf}
\usepackage[british]{babel}
\usepackage[autostyle]{csquotes}

\usepackage{algorithm2e}

\RestyleAlgo{ruled}
\SetKwComment{Comment}{/* }{ */}

\usepackage[textsize=normalsize,textwidth=2cm,colorinlistoftodos]{todonotes}

\usepackage{ctable}

\newcommand{\I}{\ensuremath{\mathrm{i}}}
\newcommand{\E}{\ensuremath{\mathrm{e}}}

\DeclareMathOperator{\imag}{Im}

\newlength{\subcolumnwidth}

\newcommand{\nextsubcolumn}[1][]{%
  \cr\noalign{\hfill}
  \if\relax\detokenize{#1}\relax\else\hsize=#1\setlength{\subcolumnwidth}{\hsize}\fi
}

\usepackage[textsize=normalsize,textwidth=2cm,colorinlistoftodos]{todonotes}

\usepackage{tikz,pgfplots}
\usetikzlibrary{arrows.meta,decorations.pathmorphing,
     decorations.markings,shadows,calc}

\pgfplotsset{compat=newest,
      width=0.9\textwidth,height=0.9\textwidth/1.618,
      every tick/.append style={black,line width=1pt},
      every axis/.append style={line width=1pt},
      enlargelimits=false,
      axis lines=middle,
      every inner x axis line/.append style={->},
      every inner y axis line/.append style={->},
      every axis y label/.style={at={(0,1)},above right},
      every axis x label/.style={at={(1,0)},above right},
      every pin edge/.style={solid,black}
}

\newcommand{\ensem}[1]{\langle #1 \rangle}

\providecommand*{\mrm}[1]{\mathrm{#1}}
\providecommand*{\eu}{\ensuremath{\mrm{e}}}
\providecommand*{\iu}{\ensuremath{\mrm{i}}}

\providecommand{\renewoperator}[3]{%
       \renewcommand*{#1}{\mathop{#2}#3}}

\renewoperator{\Re}%
               {\mathrm{Re}}{\nolimits}
\renewoperator{\Im}%
               {\mathrm{Im}}{\nolimits}

\newif\ifgreek
\def\testgreek#1{
  \ifx#1\alpha\greektrue\else
  \ifx#1\beta\greektrue\else
  \ifx#1\gamma\greektrue\else\ifx#1\Gamma\greektrue\else
  \ifx#1\delta\greektrue\else\ifx#1\Delta\greektrue\else
  \ifx#1\epsilon\greektrue\else
  \ifx#1\zeta\greektrue\else
  \ifx#1\eta\greektrue\else
  \ifx#1\theta\greektrue\else\ifx#1\Theta\greektrue\else
  \ifx#1\iota\greektrue\else
  \ifx#1\kappa\greektrue\else
  \ifx#1\lambda\greektrue\else\ifx#1\Lambda\greektrue\else
  \ifx#1\mu\greektrue\else
  \ifx#1\nu\greektrue\else
  \ifx#1\xi\greektrue\else\ifx#1\Xi\greektrue\else
  \ifx#1\pi\greektrue\else\ifx#1\Pi\greektrue\else
  \ifx#1\rho\greektrue\else
  \ifx#1\sigma\greektrue\else\ifx#1\Sigma\greektrue\else
  \ifx#1\tau\greektrue\else
  \ifx#1\upsilon\greektrue\else\ifx#1\Upsilon\greektrue\else
  \ifx#1\phi\greektrue\else\ifx#1\Phi\greektrue\else
  \ifx#1\chi\greektrue\else
  \ifx#1\psi\greektrue\else\ifx#1\Psi\greektrue\else
  \ifx#1\omega\greektrue\else\ifx#1\Omega\greektrue\else
  \ifx#1\varepsilon\greektrue\else
  \ifx#1\vartheta\greektrue\else
  \ifx#1\varrho\greektrue\else
  \ifx#1\varsigma\greektrue\else
  \ifx#1\varphi\greektrue\else
     \greekfalse
  \fi\fi\fi\fi\fi\fi\fi\fi\fi\fi
  \fi\fi\fi\fi\fi\fi\fi\fi\fi\fi
  \fi\fi\fi\fi\fi\fi\fi\fi\fi\fi
  \fi\fi\fi\fi\fi\fi\fi\fi\fi}

\renewcommand{\vec}[1]{\boldsymbol#1}

\newcommand{\particle}{\mathcal P} 
\newcommand{\reg}{\mathcal R} 
\newcommand{\numdensity}{{\mathfrak n}} 

\newcommand{\dv}{\vec{d}}
\newcommand{\rv}{\vec{r}}

\newcommand{\J}{\mathrm J}
\renewcommand{\H}{\mathrm H}
\newcommand{\uin}{u_{\mathrm{in}}}
\newcommand{\ui}{u_{\mathrm{inc}}}
\newcommand{\us}{u_{\mathrm{sc}}}

\usepackage[most]{tcolorbox} 

\newtcolorbox[auto counter]{optionalnote}[2][]{
    parbox=false,
    colbacktitle= white,
    colback=green!5!white,
    colframe=white!45!black,
    coltitle=black,
    enhanced,
    attach boxed title to top left={yshift=-1mm},
    title={\thetcbcounter.~#2}
,#1}

\newtcolorbox{highlight-result}[1][]{
 parbox=false,
 boxrule=0pt,top=0pt,bottom=0pt,
colback=blue!8!white,
enhanced,#1}

\tcbset{colback=blue!8!white, colframe=white!50!white,top=0mm,bottom=0mm,right=0mm,left=0mm}

\definecolor{Amaranth}{rgb}{0.9, 0.17, 0.31}
\definecolor{Almond}{rgb}{0.94, 0.87, 0.8}
\definecolor{Apricot}{rgb}{0.98, 0.81, 0.69}
\definecolor{Fuchsia}{rgb}{0.57, 0.36, 0.51}
\definecolor{Amethyst}{rgb}{0.6, 0.4, 0.8}

\makeatother

\graphicspath{{images/}}

\begin{document}

\title{The average transmitted wave in random particulate materials}
\author{Aris Karnezis, Paulo S. Piva, Art L. Gower}
\date{\today}

\tableofcontents
\maketitle

\begin{abstract}
Microwave remote sensing is significantly altered when passing through clouds or dense ice. This phenomenon isn't unique to microwaves; for instance, ultrasound is also disrupted when traversing through heterogeneous tissues. Understanding the average transmission in particle-filled environments is central to improve data extraction or even to create materials that can selectively block or absorb certain wave frequencies. Most methods that calculate the average transmitted field assume that it satisfies a wave equation with a complex effective wavenumber. However, recent theoretical work has predicted more than one effective wave propagating even in a material which is statistical isotropic and for scalar waves. In this work we provide the first clear evidence of these predicted multiple effective waves by using high fidelity Monte-Carlo simulations that do not make any statistical assumptions. To achieve this, we also had to fill in a missing link in the theory for particulate materials: we prove that the incident wave does not propagate within the material, which is usually taken as an assumption called the Ewald-Oseen extinction theorem. By proving this we conclude that the extinction length - the distance it takes for the incident wave to be extinct - is equal to the correlation length between the particles.
\end{abstract}

\pagestyle{plain}

\section{Introduction} 
\label{sec: intro}

Most materials, at some length scale, are formed of a random configuration of smaller particles. 
Consider particles in powder for pharmaceuticals, grains in the sand, oil droplets in emulsions, and aggregates in solid composites. The wide number of these particulate materials, and engineering applications, make it worth while to develop methods to measure these materials and design them intelligently. 

\textbf{Background.} When it comes to measurement and characterisation, the main tools are classical waves such as electromagnetic and ultrasonic or acoustic. The governing equations for these classical waves would be well understood if the material itself was known in all its details. Unfortunately, in most cases it is impossible to know in detail the microstructure of the material because it is disordered. In these scenarios, ensemble averaging and statistical assumptions need to be used to obtain solvable systems \cite{Ishimaru1978book, mishchenko2014electromagnetic, mishchenko_multiple_2006, Uscinski1977book}. 
 
How classical waves interact with particulate materials (on average) is well understood within certain limits. In the long wavelength limit, where the particles appear small compared to the wavelength of the incident wave, it is well understood how to calculate effective properties \cite{Salvatore2002}. In the dilute limit, where there is no multiple scattering, Mie theory has led to characterisation methods such as Dynamic Light Scattering and laser diffraction have led to a range of widely used tools\footnote{Malvern Panalytical: \href{https://www.malvernpanalytical.com/en/products/technology/light-scattering/laser-diffraction}{www.malvernpanalytical.com/laser-diffraction}}\footnote{Horiba Scientific: \href{https://www.horiba.com/int/scientific/technologies/dynamic-light-scattering-dls-particle-size-distribution-analysis/dynamic-light-scattering-dls-particle-size-distribution-analysis/}{www.horiba.com/dynamic-light-scattering}}.

\textbf{Pushing the limits.} In the cases where multiple scattering is significant, and the incident wavelength is not long (compared to the microstructure), the average wave is not as simple to describe \cite{Carminati2021, gower2021effective, Ma&Varadan1984}. This is especially true when using exotic pair-correlations \cite{Torquato2015, Torquato2018} and resonant particles \cite{Ma2016, Garcia2021}. To push the theory to these new limits, we need to clearly understand the validity of all the assumptions made. Within this context, we aim to address two significant assumptions that currently remain unanswered.

\textbf{Multiple effective wavenumbers.}
Most of the literature assumes there is only one effective wavenumber \cite{Dubois2011, linton_multiple_2005, PMartin2006, Tishkovets2011, Tsang1982, Varadan1983}. As the medium is isotropic and homogeneous (after ensemble averaging) it seems reasonable to assume that there is only one effective wavenumber $k_\star$ for waves travelling in a bulk material (i.e. no waveguide). However, two different theoretical methods \cite{gower2021effective,JRWillis2020} have predicted that there exist at least two (complex) effective wavenumbers for one fixed frequency. Here we give the first clear numerical evidence of these multiple effective wavenumbers as well as demonstrate that multiple wavenumbers are trigger by particles, and frequencies, that lead to strong scattering.

\textbf{Incident wave extinction.} The second assumption we address is about the presence of the incident wave inside the particulate material. It is often assumed that the incident wave does not propagate, or contribute to the total field, inside the particulate material. This assumption is called the Ewald-Oseen extinction theorem \cite{Carminati2021, Fearn1996, Lax1952, mishchenko2014electromagnetic,Tishkovets2011,Tsang1982,Tsang2000}. We address this by providing a proof that for any material geometry, frequency, and incident wave, the Ewald-Oseen extinction theorem holds when deep enough within the material. Specifically, we show that the incident wave does not propagate further than the correlation length between the particles. That is, the extinction length is equal to the inter particle correlation length.   

\textbf{Monte-Carlo.} There have been several studies that use Monte-Carlo methods to validate effective wave theory. Examples include comparing Monte-Carlo with: the average scattering from a sphere filled with particles \cite{Muinonen2012, Zurk1995},  and one effective wavenumber from the theory \cite{chekroun_time-domain_2012,chekroun_comparison_2009}. To our knowledge, there has been no Monte-Carlo validation or evidence that more than one multiple effective wavenumber exist. Here, by using precise Monte-Carlo simulations, we provide the first clear evidence that at least two effective wavenumbers, and therefore two effective waves, are present in the transmitted field. 
The theoretical methods that predict these multiple effective wavenumbers make several statistical assumptions, whereas our numerical simulations make no such assumptions. They therefore provide a clear validation of the theoretical predictions.  

\textbf{Summary of the paper.} Below in \Cref{sec:overview} we provide an overview of the theory and the results of this paper.

In \Cref{sec:plate with particles} we discuss what the theory predicts for the average wave in a plate filled with random particles, followed by how to easily identify when multiple wavenumbers should appear. 

In \Cref{sec:Monte-Carlo} we discuss our Monte-Carlo simulations, which involve simulating waves scattered from tens of thousands of particle configurations, how we verify convergence, and how we clearly demonstrate that there are scenarios where at least two effective wavenumbers appear in the transmitted field. 

In \Cref{sec:deducing transmission} we provide rigorous derivations that: the incident wave does not propagate in the particulate material, and the average transmitted wave is a sum of waves which satisfy effective wave equations. Our derivations are more general than just for a plate, they hold for a finite region, and in fact for any spatial dimension. So the proof we provide is also valid for three dimensional materials. 

\subsection{Overview of the theory} \label{sec:overview}

Consider a harmonic incident plane-wave $u_\text{inc}(x) = \eu^{\iu k x}$, satisfying Helmholtz equation with wavenumber $k$, so that $u_\text{inc}(x) \eu^{- \iu \omega t}$ satisfies a scalar wave equation, where $x$ is the distance of propagation. When this incident wave propagates through a random particulate medium, the general assumption has been that it will be replaced by one effective wave of the form:
\begin{equation} \label{eqn:average_wave_1}
    \braket{u(x)} = A_{\star} \eu^{\iu k_{\star} x},
\end{equation}
where $k_\star$ is a complex effective wavenumber, for the fixed frequency $\omega$, $A_{\star}$ is the amplitude, and $\ensem{u(x)}$ is the ensemble average of $u(x)$ over all possible particle configurations \cite{Dubois2011, linton_multiple_2005, PMartin2006, Tishkovets2011, Tsang1982, Varadan1983}. 

In the low frequency limit, when the particles are small relative to the incident wavelength, there is substantial evidence to justify~\eqref{eqn:average_wave_1}. But beyond the low frequency limit, and when using more exotic pair-correlations and distributions for the particles, there is no clear consensus. Two different methods \cite{gower2019multiple,JRWillis2020} suggest a different form for~\eqref{eqn:average_wave_1} given by
\begin{equation}\label{eqn:transmit split intro}
  \ensem{u(x)} = A_\text{inc}\eu^{\iu k x} + \sum_{p = 1}^ \infty A_p\eu^{\iu k_p x},  
\end{equation}
where the $k_p$ are complex effective wavenumbers. In the next two paragraphs we further explain the form~\eqref{eqn:transmit split}.

{\textbf{Incident wave extinction.}} When $A_\text{inc}$ is not zero, a part of the incident wave remains even when propagating through a random particulate. The work by Martin~\cite{martin2011multiple} suggests that this is the case. Assuming $A_\text{inc} = 0$ is called the Ewald-Oseen extinction theorem \cite{Fearn1996, Lax1952, Tishkovets2011}, and it is due to the scattering and absorption of energy by the random distribution of particles. Much of the literature \cite{Ballenegger1999} assumes this extinction happens after the incident wave has propagated a certain distance in the material. In this paper we prove that for any: frequency, material geometry, particle distribution, and for two and three spatial dimensions, the incident wave is extinct ($A_\text{inc} = 0$ for plane-waves) at a distance equal to particle correlation length.

{\textbf{Multiple effective wavenumbers.}} Each term in the sum of \eqref{eqn:transmit split intro} represents an effective wave with a different wavenumber $k_p$. This is highly unusual for scalar waves at a fixed angular frequency $\omega$. Yet, two different theoretical methods have predicted the existence of at least two ($p > 1$) complex effective wavenumbers \cite{gower2019multiple, Simon2021,JRWillis2020}. The evidence for this unusual prediction, beyond just theoretical, is lacking. In this work we use highly accurate simulations and Monte-Carlo method for circular cylindrical particles to demonstrate that these extra wavenumbers are present for specific frequencies. To our knowledge, this is the first clear evidence of the existence of these multiple effective wavenumbers. 

{\textbf{Average reflected field.}} Beyond just a curiosity, these extra effective wavenumbers can have a significant effect on the average reflected, transmitted, or scattered wave from a particulate material \cite{gower2019proof,gower2019multiple}. It is also far simpler to calculate the cases where there is only one dominate wavenumber~\cite{Dubois2011, gower2021effective}. 

\begin{figure}[H]
    \centering
    \includegraphics[width=0.8\linewidth]{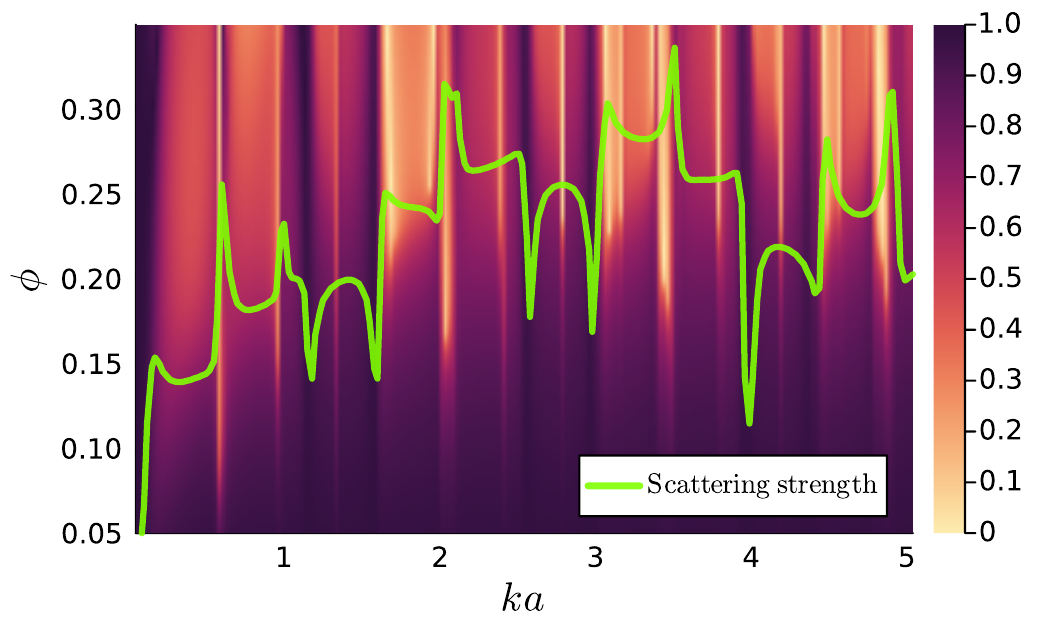}
    \caption{The phase diagram for sound-soft ($\rho_o, c_o = 0.30$) particles showing when more than one effective wavenumber is needed. The $x$--axis shows $ka$, with $a$ being the particle radius, and $k$ being the incident wavenumber. The $y$--axis is the particle volume fraction. The regions with a light colour in the background, anything less than approximately $0.5$ shown in the colour bar, require more than one effective wavenumber. The height of the green curve is the scattering strength of just one particle (given by \eqref{eqn:scattering strength}). Clearly strong scatterers lead to multiple effective wavenumbers. }
    \label{fig:soft-phase-diagram}
\end{figure}

\newpage

To understand when these multiple wavenumbers are needed, we have produced a series of phase diagrams. An example is shown in \Cref{fig:soft-phase-diagram} for sound soft particles. The regions with a light colour, anything below 0.5 shown in the colour bar, require more than one effective wavenumber to accurately describe the transmitted field. For more details see \Cref{fig: phase_diagrams} and the discussion around that figure. For example, we can see that only one wavenumber is needed for low particle volume fraction $\phi < 0.1$, with only one exception around $ka = 0.6$. The green curve in the figure shows the scattering strength of just one particle by itself. Surprisingly, we see that at the frequencies at which the single particle scatters the strongest (the peaks in the green curve) is also the frequencies at which two (or more) effective wavenumbers are required, and therefore needed to accurately describe wave transmission. These results, and other phase diagrams, are further discussed in Section \ref{subsec: MonteCarlo}.


\section{A plate filled with particles}
\label{sec:plate with particles}

For a medium that is isotropic and homogeneous, it seems reasonable to assume that there is only one effective wavenumber for one fixed angular frequency $\omega$. However, recently two different theoretical models have predicted that there exist at least two (complex) effective wavenumbers \cite{gower2019multiple,JRWillis2020}, as shown in \eqref{eqn:transmit split}. 

Here we design a computational experiment to give clear evidence of at least two of these effective wavenumbers. We do this by using a robust numerical method based on high fidelity Monte-Carlo simulations for two dimensional disks.

\begin{figure}[H]
  \centering
  \includegraphics[height= 0.5\linewidth]{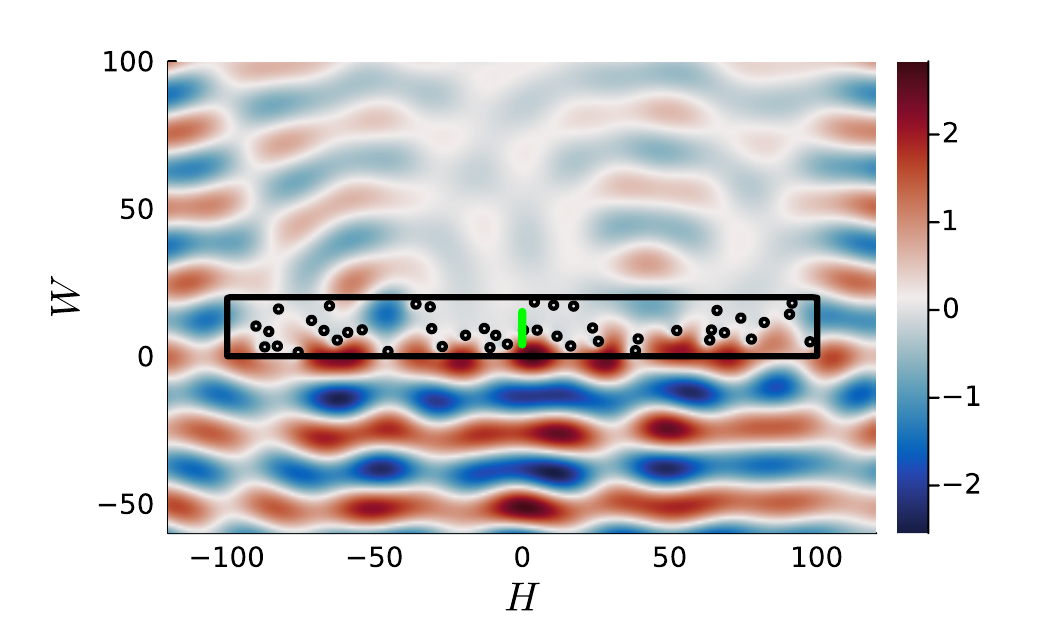}
  \caption{Scattering of an incident plane-wave approaching from the bottom, onto one specific configuration of randomly distributed circular cylinders (or particles) $\Lambda$. The simulation directly solves the governing equations \cite{martin_multiple_2006}, and the green line shows where the field is measured. }
  \label{fig:plate_geometry}
\end{figure}

To describe the material, let $\particle_j$ be the disk occupied by the $j$-th particle, as represented by the circles in \Cref{fig:plate_geometry} and shown in more detail in \Cref{fig:regions}. Let $\particle = \cup_j \particle_j$ be the union of all particles. For simplicity we consider circular particles of equal size. In other words, using standard set-builder notation
\begin{equation}
  \particle_j = \{\rv \in \mathbb R^2 : \; |\rv - \rv_j| < a\},
\end{equation}
where $a$ is the radius of the particle, and $|\vec x|$ is the length of the vector $\vec x$.

The particles are restricted inside the plate geometry $\mathcal R$, given by
\begin{equation}
    \label{eq:plate_definition}
    \mathcal{R} = \left\{(x, y) \in \mathbb R^2 : 0 \leq x \leq W, \ -\dfrac{H}{2} \leq y \leq \dfrac{H}{2} \right\},
\end{equation}
where $W$ is the width and $H$ is the height of the plate as shown in \Cref{fig:plate_geometry}.

The total field $u(\rv)$ satisfies a Helmholtz equation which depends on whether $\rv$ is inside a particle or not:
\begin{align} 
    \label{eq: scalar_wave}
    \nabla^ 2 u(\rv) + k^2 u(\rv) &=  0, \qquad \text{for} \;\; \rv \in \mathcal R \setminus \particle,
    \\
    \nabla^ 2 u(\rv) + k_{0}^2 u(\rv) &=  0, \qquad \text{for} \;\;\rv \in \particle,
\end{align}
where $k = {\omega}/{c}$ and $k_0 = {\omega}/{c_0}$ are the real wavenumbers of the background and particles respectively. The scalars $c$ and $c_0$ are, respectively, the wavespeeds in the background and particles.

The simplest scenario to numerically check for effective waves is for planar symmetry. As in this case each frequency has only one mode: the plane-wave.  To achieve this, we fill a plate region with a configuration of randomly distributed cylindrical particles, as shown in \Cref{fig:plate_geometry}. The particles are identical, except for their positions. 

\subsection{Effective waves for planar symmetry} \label{sec:effective-planar}

Here we summarise the results of the theory for plane-wave symmetry. The results here will be compared with a Monte-Carlo method detailed in the next section.

We consider an incident plane-wave of the form:
\begin{equation}\label{eqn:incident-plane-wave}
    \ui(x) = \eu^{\iu k x},
\end{equation}
and consider particles in a plate region $\mathcal R$ with an infinite height, whereas \Cref{fig:plate_geometry} shows a truncated plate with a finite height. 

The theoretical methods consider an ensemble average of the total field $u$. To achieve this, we describe one configuration of particles with
\begin{equation}
\Lambda = (\rv_1,\rv_2,\ldots, \rv_J),    
\end{equation}
where $\rv_j$ is the centre position of the particle $\particle_j$. Naturally, the field $u$ depends on the particle positions. To make this explicit we use $u(x;\Lambda)$.

Next, to calculate the ensemble average, we need to define the probability of all possible particle configurations. To do this, we introduce the joint probability density given by $p(\Lambda)$. For a brief overview on the probability density function $p$, see \cite{gower_reflection_2018, gower2021effective, linton_multiple_2005}. The theoretical methods then calculate and predict the ensemble average defined by
\begin{equation} \label{def:ensemble average}
    \ensem{u(x)} := \int u(x;\Lambda) p(\Lambda) \mathrm d \Lambda,
\end{equation}
where the integral is over all possible particle positions, and the fields depend only on the spatial position $x$ as we are considering planar symmetry.

It is widely assumed that the average $\ensem{u(x)}$ satisfies a Helmholtz equation with a unique effective complex wavenumber $k_*$ \cite{Dubois2011, linton_multiple_2005, PMartin2006, Tishkovets2011, Tsang1982, Varadan1983}. In \Cref{sec:trans effective} we prove that $\ensem{u(x)}$ is a sum of several effective waves, but only when $x$ is deep enough within the material. To define what ``deep enough'' means we need to introduce the particle pair-correlation.

\textbf{Pair-correlations.} We assume that particles are distributed both homogeneously and isotropically, which leads to
\begin{equation} \label{eqn:pair-corr-to-p}
    g(|\rv_1 - \rv_2|) = \frac{p(\rv_1, \rv_2)}{p(\rv_1) p(\rv_2)},
\end{equation}
for an infinite number of particles. For details on pair-correlations see \cite{kong2004scattering}. 

\begin{figure}[H]
    \centering
    \includegraphics[width=0.6\linewidth]{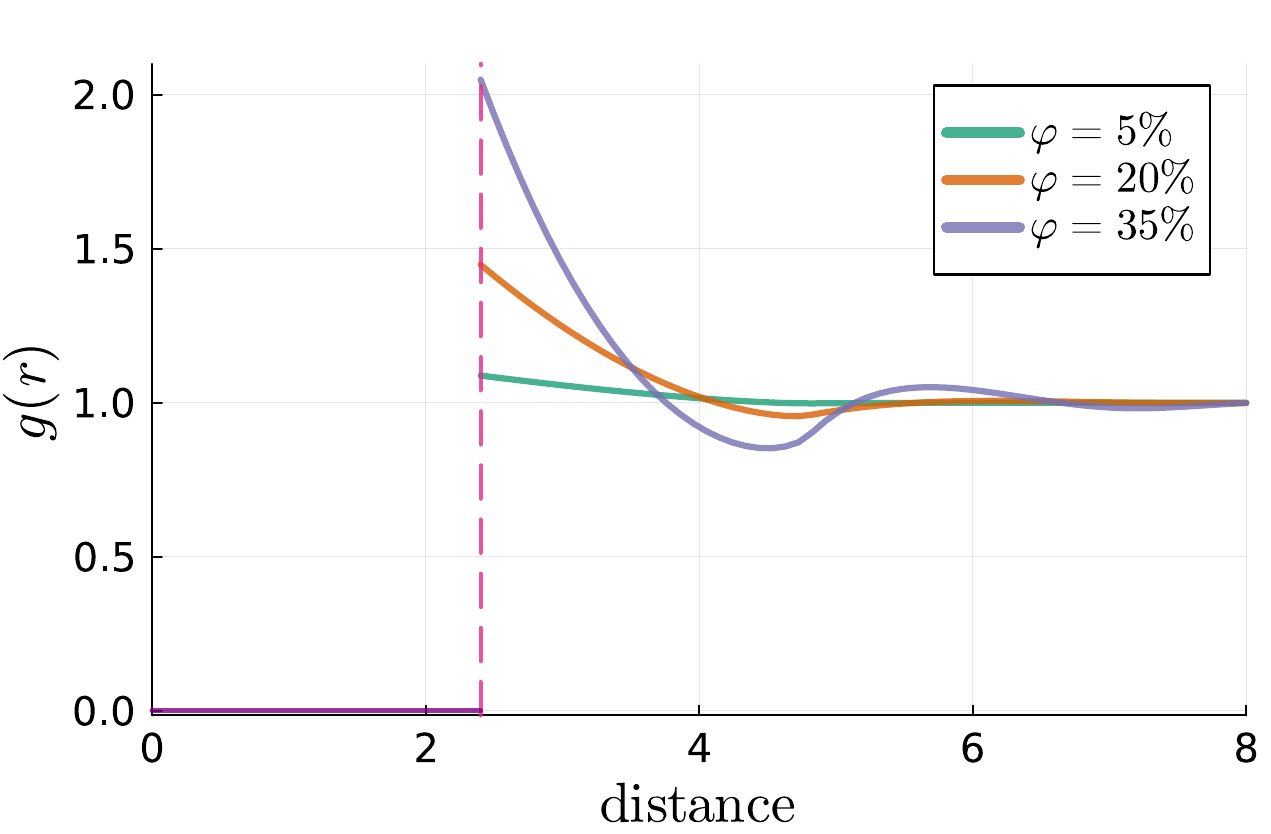}
    \caption{The Percus-Yevick approximation is a pair correlation 
    that represents particles that are uniformly randomly placed, except particles do not overlap. That is, particles do no attract or repel each other. The particle radius $a = 1.2$ and $\phi$ is the particle volume fraction. 
    }
    \label{fig:Percus-Yevick}
\end{figure}

For a disordered or random configuration of particles we have that 
\begin{equation}\label{def:pair-correlation-g}
  g(|\rv_1 - \rv_2|) =  \begin{cases}
    0  &  \text{if} \quad |\rv_1 - \rv_2| \leq a_{12},
    \\
    1  & \text{if} \quad  |\rv_1 - \rv_2| \geq b_{12},
  \end{cases}
\end{equation}
where $b_{12} \geq a_{12} > 2a$. For the region $a_{12} \leq |\rv_1 - \rv_2| \leq b_{12}$ the pair correlation can take any values for our calculation below, though we expect $g(r)$ to be continuous in $r$.   
In this work we use two different pair-correlations. The first, and the simplest, is called \emph{Hole-Correction}, which assumes that $b_{12} = a_{12}$. The second is called the \emph{Percus-Yevick} approximation, which more accurately approximates the pair correlation for particles that are distributed according to a uniform random probability, except no two particles can overlap \cite{Bringi1982, kong2004scattering, Gerhard2021, Tsang1982}. We use the results from \cite{adda2008solution} to obtain Percus-Yevick for disks. \Cref{fig:Percus-Yevick} shows the Percus-Yevick distributions for several different particle volume fractions $\phi$.    

For disordered media we have that $g(r) \to 1$ as $r \to \infty$. That is, particles become uncorrelated as they are further apart. To prove that $\ensem{u(x)}$ is a sum of effective waves we need a slightly stronger assumption: that there is a distance $b_{12}$ at which particles are completely uncorrelated as used in \eqref{def:pair-correlation-g}. 

\textbf{Effective plane-waves.} By having a length $b_{12}$ at which particles become uncorrelated, we demonstrate in \Cref{sec:trans effective} that for an infinite plate geometry ($H \to \infty$ in \eqref{eq:plate_definition}) filled with particles
\begin{equation} 
\label{eq:effective_waves}
\ensem{u(x)} = \sum_{p=1}^{P} \left( A_p^+ \eu^{\iu k_p x} + A_p^- \eu^{-\iu k_p x} \right) \quad \text{for} \;\;\; \left| x - \frac{W}{2} \right| < \frac{W}{2} - b_{12} - a, 
\end{equation}
where we used planar symmetry as shown in \cite{gower2021effective}. The $A_p^{\pm}$ are complex amplitudes, the $k_p$ are the complex effective wavenumber, and $P$ is the number of effective wavenumbers. There is an infinite number of wavenumbers $P$, but according to theoretical calculations only a few are needed for accurate results. For a detailed discussion on these multiple effective wavenumbers see \cite{gower2019proof, gower2019multiple}. 

\textbf{The dispersion equation.} To calculate the wavenumbers $k_p$ we use the dispersion equation appearing in \cite{gower2019proof, gower2019multiple, linton_multiple_2005}. The assumptions needed to arrive at this dispersion equation are shown in \Cref{sec:deducing transmission}. To summarise, the $k_p$ are determined by solving
\begin{equation} \label{eqn:planewave-dispersion}
    \det \vec M(k_\star) = 0, 
    \;\;\; \text{with} \;\;\;
    M_{nn'}(k_\star) = \delta_{n n'}
    +2\pi \mathfrak{n} T_n 
    [\mathcal N_{n'-n} - \mathcal G_{n'-n}],
\end{equation}
where
\begin{align}
    & \mathcal N_\ell
    = \frac{1}{k_\star^2 - k^2 } (k a_{12}\mrm H'_{l}(k a_{12})\mrm J_{l}(k_\star a_{12}) - k_\star a_{12}\mrm H_{l}(k a_{12})\mrm J'_{l}(k_\star a_{12})),
    \\
    & \mathcal G_\ell =  \int_{a_{12}}^{b_{12}}\mathrm J_{\ell}(k_\star r)\mathrm H_{\ell}(kr)(g(r) - 1)r\mathrm dr,
\end{align}
and the term $\mathfrak n$ is the average number of particles per area, $r = |\rv|$, $ \delta_{n n'}$ is the Kronecker delta, and $\mrm H_\ell$ is the Hankel function of the first kind, while $\mrm J_\ell$ is the Bessel function. The term $T_n$ is the T-matrix which determines how one particle scatters waves by itself. For circular homogeneous particles in acoustics we have that
\begin{equation} \label{eqn:circular_t-matrix}
    T_{n} = - \frac{\gamma \J_n' (k a) \J_n (k_o a) - \J_n (k a) \J_n' (k_o a)}
                   {\gamma \H_n'(k a_o) \J_n(k_o a) - \H_n(k a) \J_n'(k_o a)},
  \end{equation}
where $\gamma = {\rho_o c_o}/{\rho c}$ and $k_o = {\omega}/{c_o}$, with $\rho_o$ being the mass density of the particles and $c_o$ being the wavespeed within the particles. 

There are infinitely many $k_\star$ which solve $\det \vec M(k_\star) = 0$.  We denote  these solutions as $k_1, \,k_2, \ldots$. 
The main objective of our Monte-Carlo simulations is to check if the theoretical predictions of the wavenumbers $k_p$ are accurate, and to clearly demonstrate that there is more than one effective wavenumber appearing in the Monte-Carlo results. Before doing this, let us first explore the effective wavenumbers predicted by solving the dispersion equation~\eqref{eqn:planewave-dispersion}. 

\textbf{Effective wavenumbers.} We want to identify when the dispersion equation \eqref{eqn:planewave-dispersion} predicts that there is more than one effective wavenumber that has a significant contribution to the average transmitted wave. It is important to understand when this occurs, as it is far simpler to calculate the average field when there is only one effective wavenumber \cite{gower2019multiple, gower2021effective}. 

Most of the scientific community at present is also not aware that more than one effective wavenumber can be excited~\cite{Carminati2021, martin_multiple_2006, mishchenko2014electromagnetic}, for just one scalar wave. So we will identify for which parameters we can run a heavy Monte-Carlo simulation, as detailed in the next section, to find clear numerical evidence of multiple effective wavenumbers.

\begin{figure}[H]
    \centering
    \includegraphics[width=0.98\linewidth]{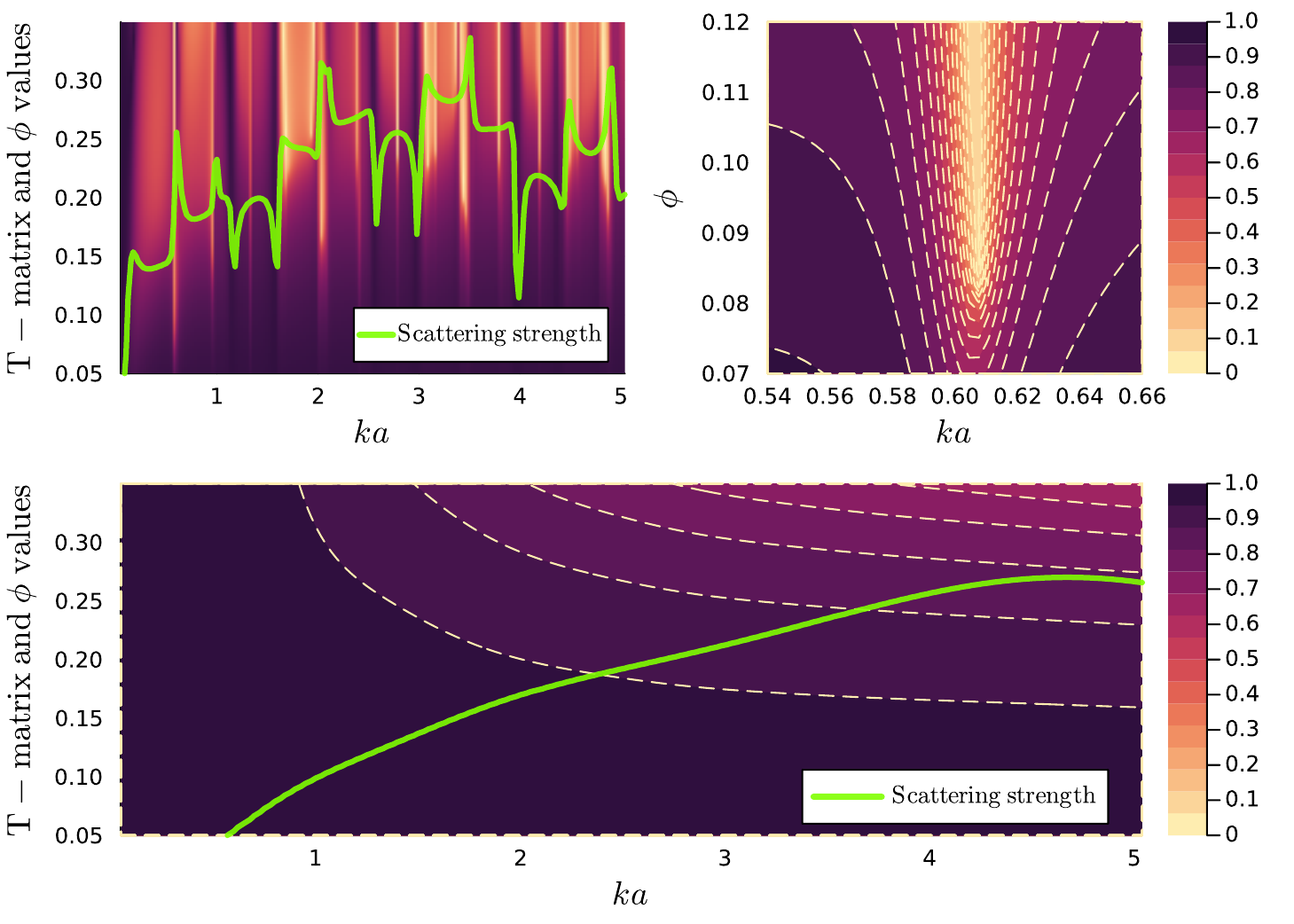}
    \caption{The phase diagrams showing when more than one effective wavenumber is needed. The top two diagrams are for sound-soft particles and the bottom diagram is for sound-hard particles with properties shown in \Cref{table:particle properties}. The colour is given by \eqref{eq:phase_diagram_measure} where the lighter colours (those above 0.5 shown in the color bar) indicate that more than one effective wavenumber can be excited. 
    The green curve shows the scattering strength of just one particle and is given by \eqref{eqn:scattering strength}.
    }
    \label{fig: phase_diagrams}
\end{figure}



\begin{table}[H] 
\centering 
    \begin{tabular}{c c c}
        Sound-soft particles & $\rho_o = 0.30 \ \mathrm{kg} / \mathrm m^{2}$ & $c_o = 0.30 \ \mathrm m / \mathrm s$ \\
        \specialrule{0.03em}{0.1em}{0.1em}
        Sound-hard particles & $\rho_o = 10.0 \ \mathrm{kg} / \mathrm m^{2}$ & $c_o = 10.0 \ \mathrm m / \mathrm s$ \\
        \specialrule{0.03em}{0.1em}{0.1em}
    \end{tabular}
    \caption{\label{table:particle properties} Shows the two main particle properties used for the numerical results. Note that sound-soft (sound-hard) particles are strong (weak) scatterers. }
\end{table}

According to theoretical results, only one effective wavenumber $k_1$ is needed when Im $k_1 \ll$ Im $k_p$ for $p=2,3, \dots$. In any other case, more than one effective wavenumber can be excited and contribute to the average transmission \cite{gower2019proof, gower2019multiple}. Though we note that the form of the incident wave and geometry of the material also affect how the wavenumbers are excited \cite{gower2021effective}.


The first step is to sweep the parameter space by varying the frequency $\omega$ and particle volume fraction $\phi$, and for each value calculate the effective wavenumbers $k_p$ by solving~\eqref{eqn:planewave-dispersion}. We do this for both sound-hard and sound-soft particles by changing $\rho_o$ and $c_o$ in \eqref{eqn:circular_t-matrix}. The two main particle properties used are shown in \Cref{table:particle properties}. Next, based on the results shown in \cite{gower2019multiple,gower2019proof}, we can estimate where more than one effective wavenumber is excited by plotting a heatmap where the colour is given by 
\begin{equation} \label{eq:phase_diagram_measure}
    {\text{colour} = \bigg| \dfrac{\imag k_{{2}}}{\imag k_{{1}}} - 1 \bigg|}.
\end{equation}

After many failed attempts, we find that the measure \eqref{eq:phase_diagram_measure} is closely related to the scattering strength of a single particle, given by
\begin{equation} \label{eqn:scattering strength}
    \text{Scattering strength} = \sqrt{\sum_n |T_n|^2},
\end{equation}
where $T_n$ is the T-matrix given by \eqref{eqn:circular_t-matrix} for acoustics.

The results of sweeping over frequency and particle volume fraction are shown in \Cref{fig: phase_diagrams}. We call these figures phase diagrams, as we see sudden shifts from only one effective wavenumber to two or more wavenumbers.  The regions with lighter shading correspond to cases where the value of \eqref{eq:phase_diagram_measure} is low, so more than one effective wavenumber is excited. Conversely, the regions with dark shading is where only one effective wavenumber is excited. 
The green curves shown on-top of the phase diagrams are the scattering strength for just one particle. Clearly we see that a large scattering strength leads to more than one effective wavenumber, which is an important observation, as calculating \eqref{eqn:scattering strength} is far simpler than calculating the wavenumbers $k_p$.  

\begin{figure}[H]
\centering
\begin{subfigure}{.5\textwidth}
  \centering
  \includegraphics[width=\linewidth]{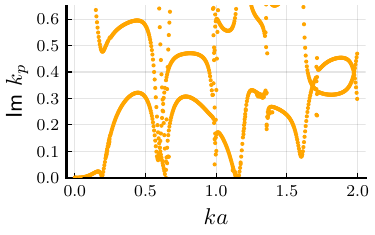}
  \caption{sound-soft scatterers}
  \label{fig:imaginary_parts_soft}
\end{subfigure}%
\begin{subfigure}{.5\textwidth}
  \centering
  \includegraphics[width=\linewidth]{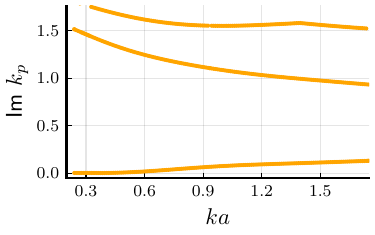}
  \caption{sound-hard scatterers}
  \label{fig:imaginary_parts_hard}
\end{subfigure}
\caption{Depicts the imaginary part of the wavenumbers $k_p$ with respect to the non-dimensional frequency $ka$. We only show the three wavenumbers with the smallest imaginary parts as these are the only ones which make a significant contribution to the average wave. The lowest curve represents $k_1$, which is the easiest to excite. The particles occupy $25\%$ of the material in each case and their properties for sound soft and hard particles are shown in Table \eqref{table:particle properties}.}
\label{fig:attenuations}
\end{figure}

The phase diagrams in \Cref{fig: phase_diagrams} show a large region of the parameter space, but to see in more detail when two or more effective wavenumbers are needed it helps to plot the imaginary parts of the wavenumbers $k_p$ against frequency, which we do in \Cref{fig:attenuations} for a particle volume fraction of $25\%$. 
For sound-soft particles (\Cref{fig:imaginary_parts_soft}), there are many frequencies $ka$ where two or even three effective wavenumbers have a similar imaginary part, meaning that these wavenumbers can be excited. In contrast, for sound-hard particles (\Cref{fig:imaginary_parts_hard}), there is only one effective wavenumber that can be easily excited, as there is one curve, representing $k_1$, that has a significantly lower imaginary part than all the others.

In the next section, we use the results presented here to identify specific scenarios where performing a Monte-Carlo simulation would be appropriate. This will allow us to verify whether the predictions of more than one effective wavenumber are indeed accurate.

\section{The Monte-Carlo simulation} \label{sec:Monte-Carlo}
A simple way to approximate the average field~\eqref{def:ensemble average} is to perform a simulation for each particle configuration and then take an average over all configurations. As we are focusing on uniformly distributed particles, the probability density $p(\Lambda)$ is a constant if particles do not overlap, and  $p(\Lambda) = 0$ if any two particles do overlap. This allows us to numerically approximate the ensemble average
\begin{equation}
    \label{eq:monte-carlo-average}
    \ensem{u(\rv;\Lambda)} = \frac{1}{S}\sum_{s=1}^S u(\rv; \Lambda_s),
\end{equation}
where each $\Lambda_s$ is one randomly sampled configuration of particles within the plate that depends on the parameter $s$.
That is, we first create a configuration of particles $\Lambda_s$, then simulate the scattered waves $u(\bm{r}; \Lambda_s)$, and then we repeat this process for $S$ configurations of particles, until the average shown above converges. As mentioned in previous sections, $\ensem{u(\rv;\Lambda)}$ should converge to a sum of plane-waves as shown by \eqref{eq:effective_waves}.

The setup for our Monte-Carlo simulations is shown in \Cref{fig:plate_geometry} (note the $x$-axis is the vertical axis in the figure), and we used an incident wave of the form $u_\text{inc}(x) = \eu^{\iu k x}$. To calculate the field $u(\rv; \Lambda_s)$ for each configuration we use a multipole expansion for each particle, together with addition translation matrices, to solve the boundary conditions \cite{martin_multiple_2006}. Specifically, we solve \cite[Equation (2.9)]{gower2019multiple} for each $s$, calculate \eqref{eq:monte-carlo-average} and evaluate the field at $\ensem{u(x, 0; \Lambda_s)}$ with the $x$ values satisfying \eqref{eq:effective_waves}. This method accurately approximates the exact solution when increasing the truncation order of the multipole expansion until reaching convergence.  

An initial numerical investigation revealed that it was computationally feasible to simulate a finite number of particles within the region $0 \leq x \leq 20$, so the width of plate $W = 20$ and height $H = 400$.  
For details on the methodology of our Monte-Carlo simulations, see \Cref{app:MC methodology}.

\subsection{The Monte-Carlo results} \label{subsec: MonteCarlo}

Based on the results of the previous section, we choose several cases to simulate the average field with a Monte-Carlo method. We want to identify cases where two or even three wavenumbers can be excited for frequencies as low as possible, because increasing the frequencies leads to a significantly larger computational cost for the Monte-Carlo simulations. Which is why we only performed Monte-Carlo simulations for three different frequencies, beyond the low frequency limit. See \Cref{app:MC methodology} for details, and descriptions on computational cost.


\begin{figure}[H]
    \begin{subfigure}{0.49\textwidth}
        \centering
        \includegraphics[width=0.9\linewidth]
        {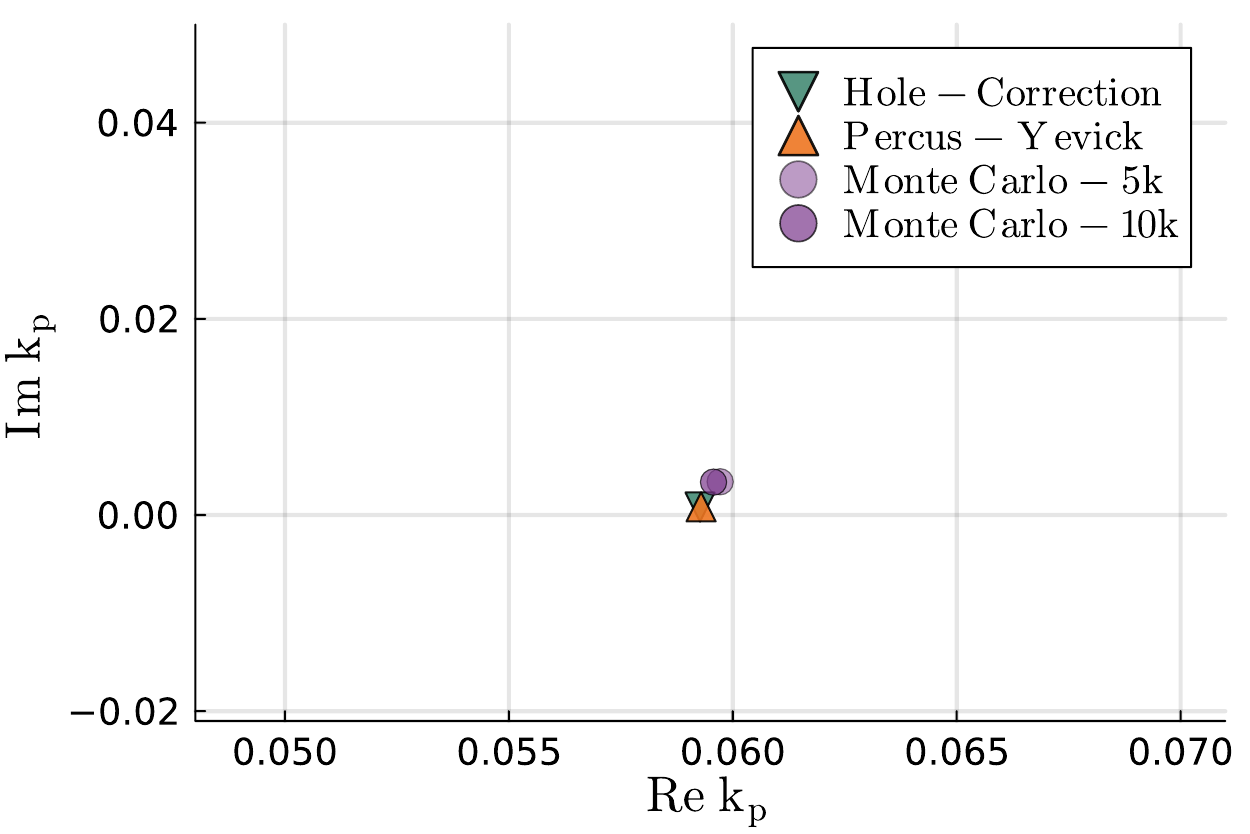}
        \caption{$\rho_0 = c_0 = 0.30, \ \phi = 5 \%, \ k a = 0.04$}
        \label{fig:low_phi_low_omega}
    \end{subfigure}
    \begin{subfigure}{.49\textwidth}
        \centering
        \includegraphics[width=0.9\linewidth]
        {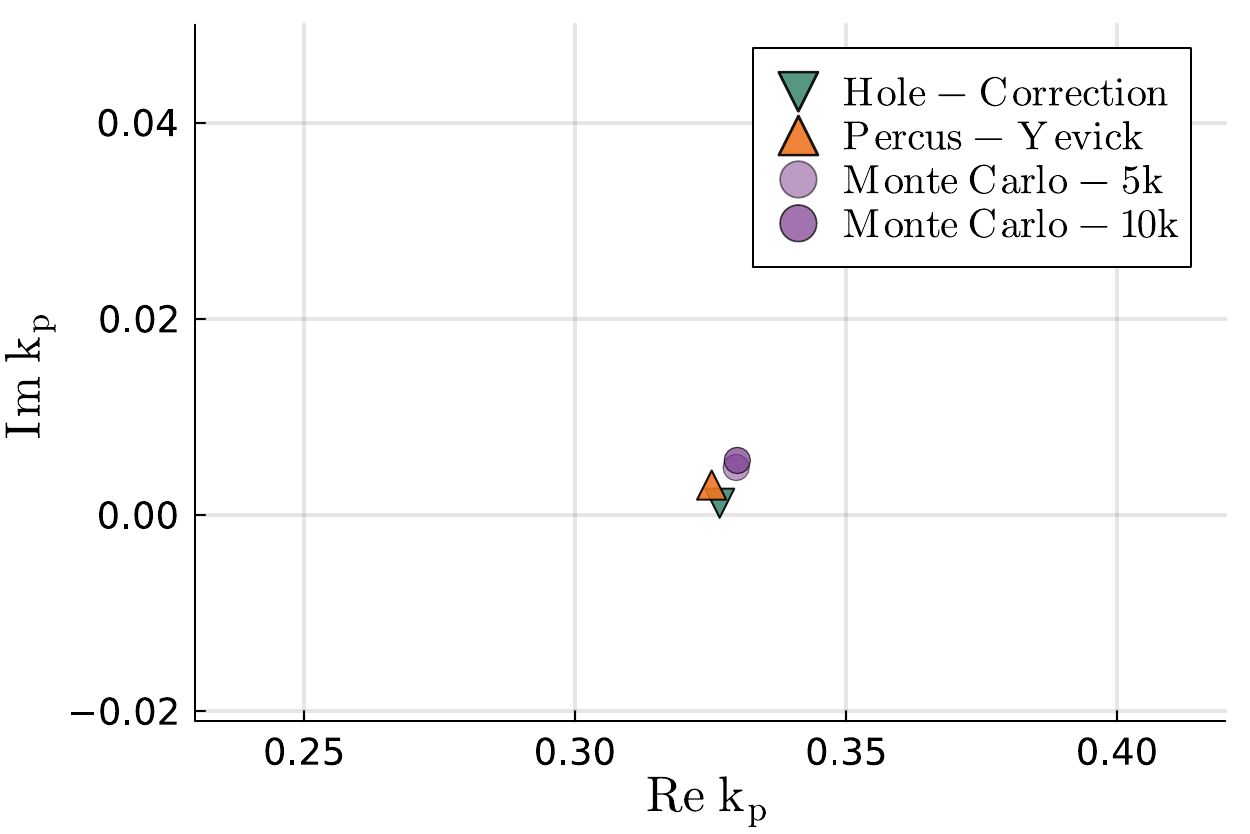}
        \caption{$\rho_0 = c_0 = 10.0, \ \phi = 25 \%, \ k a = 0.36$}
        \label{fig:1kps_hard}
    \end{subfigure}

    \caption{Shows two cases where the dispersion equation \eqref{eqn:planewave-dispersion} predicts there is only one effective wavenumber $k_p$ with a lower imaginary part, and therefore this is the only wavenumber that can be excited.
    The green and red triangles represent effective wavenumbers predicted by the \eqref{eqn:planewave-dispersion} when using either the Hole-Correction or Percus-Yevick pair-correlation. The purple dotted points represent the effective wavenumber which best fits the Monte-Carlo simulations when using the formula \eqref{eq:effective_waves}. } 
    
    \label{fig:wavenumbers_ZOOM}  
\end{figure}

\textbf{One effective wavenumber.} From \Cref{fig:imaginary_parts_hard} we see that for sound-hard particles there is a broad range of frequencies where there is only one effective wavenumber with an imaginary part which is far smaller than all others. This means that is easy to excite this wavenumber, but very difficult to excite the others. This message is also confirmed by the phase diagram shown in \Cref{fig: phase_diagrams}. To verify there is only one effective wavenumber, we perform Monte-Carlo simulations for $ka = 0.36$ for sound-hard particles, and fit the formula \eqref{eq:effective_waves}. The results shown in \Cref{fig:1kps_hard} confirm that there is only one wavenumber present in the Monte-Carlo simulations.

There is a small discrepancy between the effective wavenumbers from the dispersion equation~\eqref{eqn:planewave-dispersion} and the fitted effective wavenumbers, which could be due to errors introduced in the Monte-Carlo simulations due to truncating an infinite region.

On the other hand, for sound-soft particles we see from \Cref{fig:imaginary_parts_soft} that only for lower frequencies $0 \leq ka \leq 0.2$, only one effective wavenumber with a smaller imaginary part exists. Monte-Carlo simulations, shown in \Cref{fig:low_phi_low_omega}, confirm that there is only one effective wavenumber as predicted from the theory.



\textbf{Fitting for several effective wavenumbers.}  Increasing the frequency for sound-soft particles leads to many frequencies where two, or more, effective wavenumbers have a lower imaginary part. For example, for $0.25 < ka < 0.63$ there are two wavenumbers with a smaller imaginary part, which means it is possible to excite two effective wavenumbers in this frequency range. To exemplify, we choose two different frequencies to perform Monte-Carlo simulations: 1) $ka = 0.36$, where we aim to excite two effective wavenumbers, and 2) $ka = 0.62$ where we are aim to excite three effective wavenumbers. 

\begin{figure}[H]
\centering
\begin{subfigure}{.5\textwidth}
  \centering
  \includegraphics[width=0.9\linewidth]{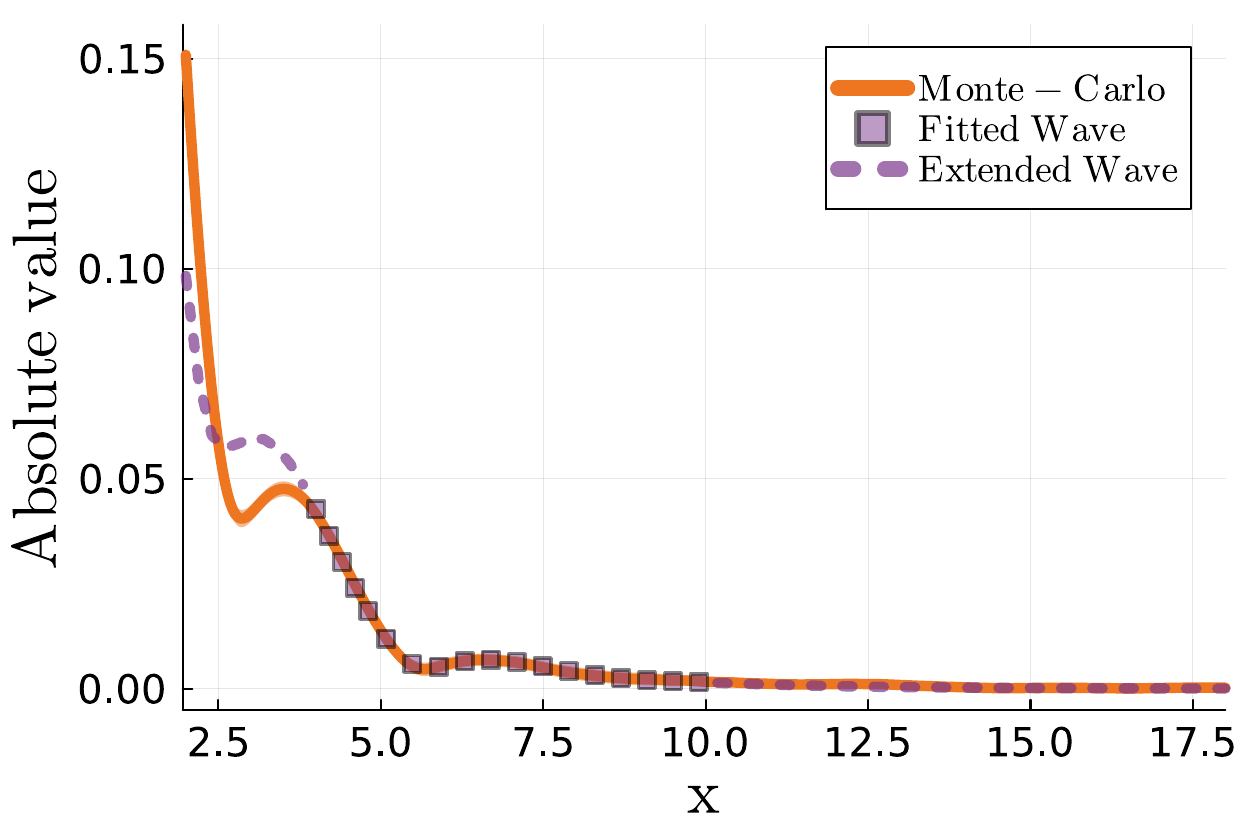}
  \caption{$\rho_0 = c_0 = 0.30, \ \phi = 25 \%, \ k a = 0.36$}
  \label{fig:average_wave_2kps}
\end{subfigure}%
\begin{subfigure}{.5\textwidth}
  \centering
  \includegraphics[width=0.9\linewidth]{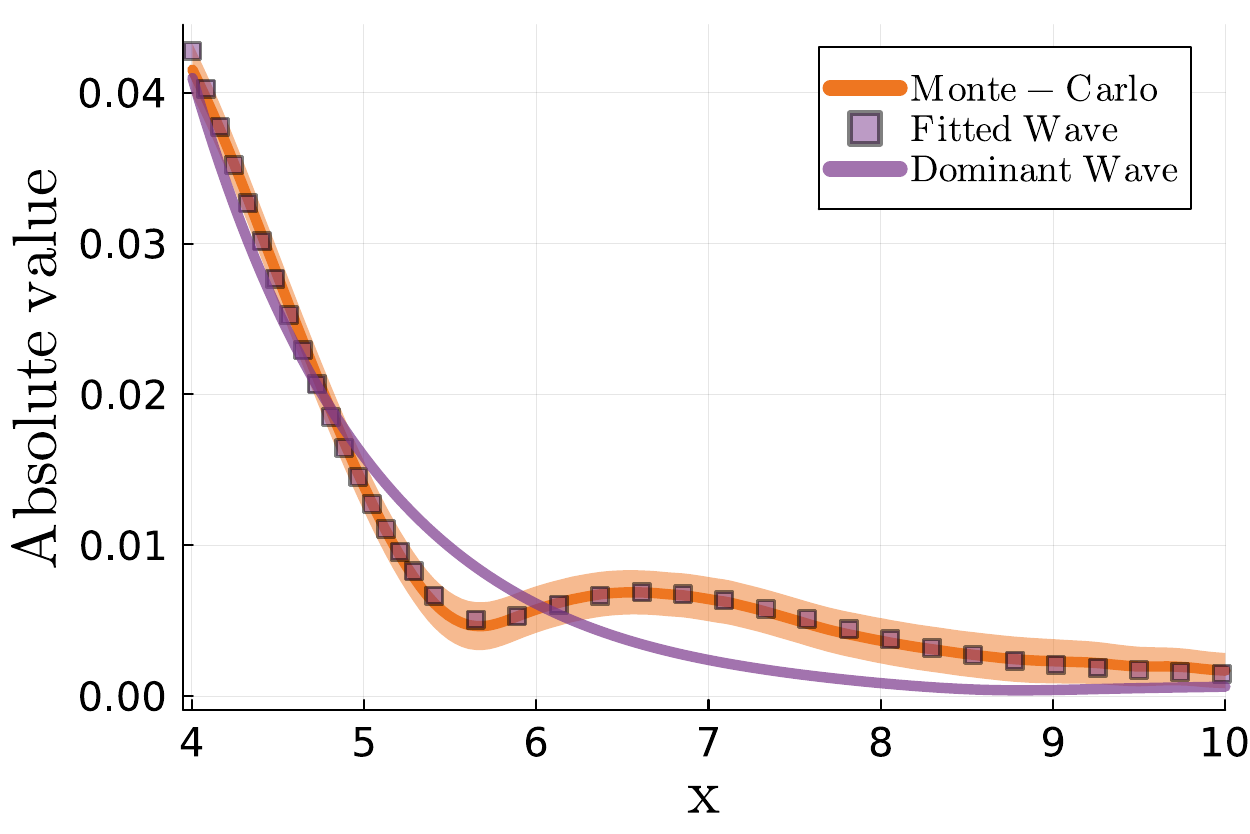}
  \caption{$\rho_0 = c_0 = 0.30, \ \phi = 25 \%, \ k a = 0.36$}
  \label{fig:average_comparison_2kps_with_dom}
\end{subfigure}
\begin{subfigure}{.5\textwidth}
  \centering
  \includegraphics[width=0.9\linewidth]{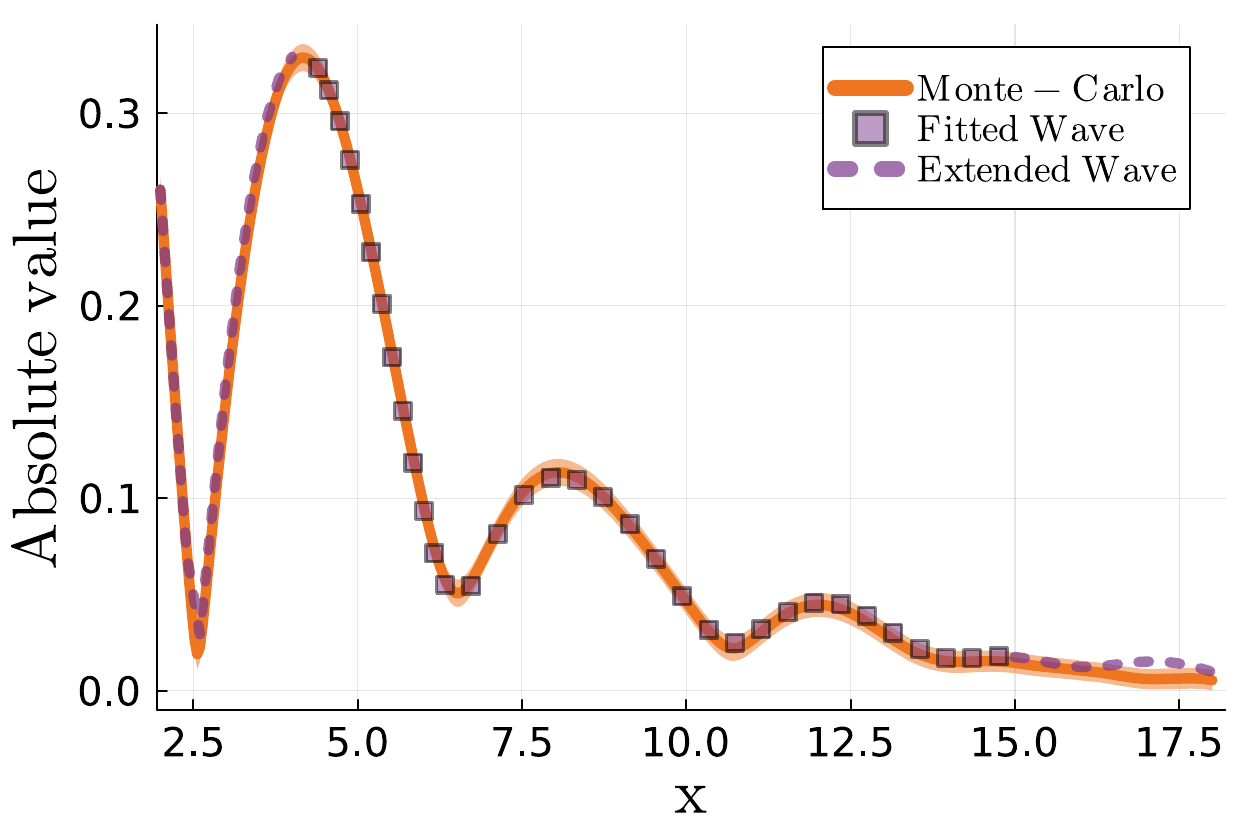}
  \caption{$\rho_0 = c_0 = 0.30, \ \phi = 25 \%, \ k a = 0.62$}
  \label{fig:average_wave_3kps}
\end{subfigure}%
\begin{subfigure}{.5\textwidth}
  \centering
  \includegraphics[width=0.9\linewidth]{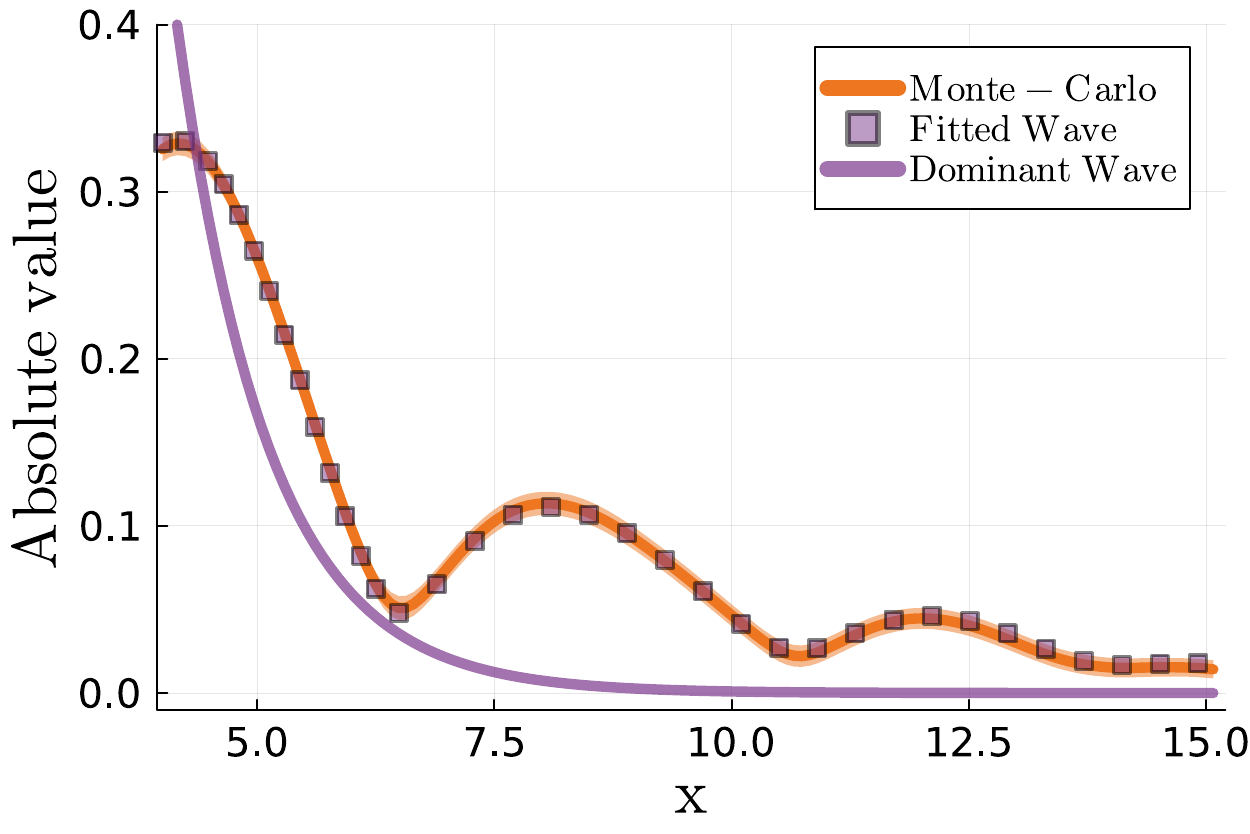}
  \caption{$\rho_0 = c_0 = 0.30, \ \phi = 25 \%, \ k a = 0.62$}
  \label{fig:average_comparison_3kps_with_dom}
\end{subfigure}
\caption{
The graphs above compare the average field from a Monte-Carlo simulation for particles in a plate, as shown in \Cref{fig:plate_geometry}, with two types of fitted waves. The frequency $ka$, particle properties and volume fraction $\phi$ used are shown below each figure.     
The \textit{Dominant Wave} is the result of fitting for just one effective wavenumber when using the formula \eqref{eq:effective_waves}, and is currently believed to be accurate by most working in the field. We see here that it is not possible to fit for just one wavenumber. The \textit{Fitted wave} is a result of fitting the formula \eqref{eq:effective_waves} with $P =2$ for the top two graphs and $P=3$ for the bottom two, where the \emph{Extended Wave} shows what the formula \eqref{eq:effective_waves} predicts outside of the fitted region. }
\label{fig:average_comparison}
\end{figure}

\Cref{fig:average_comparison} shows how the formula \eqref{eq:effective_waves} fitted to the Monte-Carlo data clearly matches the Monte-Carlo data, whereas the graphs on the right show how fitting for only one effective wavenumber, termed the ``Dominate wave'', does not match the Monte-Carlo data. The two panels on the left also show extending the fitted wave, by using the formula \eqref{eq:effective_waves}, beyond the fitted region can still closely follow the Monte-Carlo results, but is expected that as the extended field gets closer to $x = 0$, the match with the Monte-Carlo field will get worse due to a boundary layer where many effective wavenumbers become significant \cite{gower2019proof,gower2019multiple}. 
The presence of this boundary layer, and our theoretical results in \Cref{sec:deducing transmission}, have all guided how best to perform, and fit to, the Monte-Carlo results as discussed in \Cref{app:MC methodology}. 

\textbf{Avoid overfitting.}  Having an accurate fit does not necessarily give strong evidence that the formula \eqref{eq:effective_waves} is correct. This is specially true for higher frequencies where the Monte-Carlo simulation has a higher standard error of the mean, shown by the shaded yellow region. A larger standard error means that there is a range of effective wavenumbers which can fit this data, and still be closer to the Monte-Carlo results than the standard error. Further, fitting a sum of plane-waves, such as shown by \eqref{eq:effective_waves}, can lead to over-fitting when using too many wavenumbers, and even become ill-posed. We explain more details on this in the Monte-Carlo methodology in Appendix \ref{app:MC methodology}.

To verify that we avoid over-fitting, and check if small changes in the field lead to large changes in the fitted wavenumbers, we develop a method, called the projection method, which fits the formula \eqref{eq:effective_waves} for all possible combinations of the effective wavenumbers in a region. Details are given in Appendix~\ref{app:MC methodology}. 

\begin{figure}[H]
    \centering
    \includegraphics[width=0.7\linewidth]{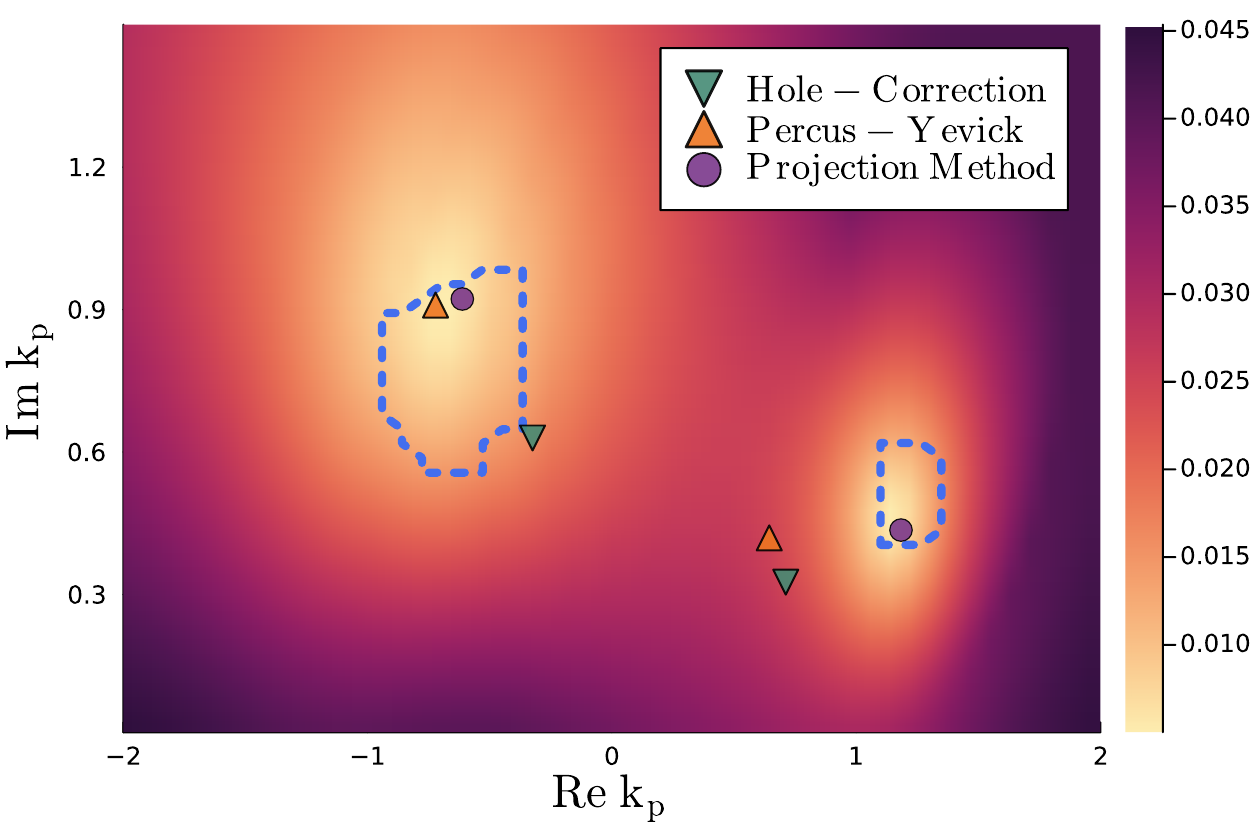}
    \caption{Shows how two wavenumbers are needed to fit the formula \eqref{eq:effective_waves} to the Monte-Carlo results, for the particle properties $\rho_0 = c_0 = 0.30, \ \phi = 25 \%$ and frequency $k a = 0.36$. When using the two best fits, shown by \emph{Projection method}, we obtain the fitting shown in \Cref{fig:average_wave_2kps}. The density plot shows what regions of complex wavenumbers that best fit the Monte-Carlo results. When using one wavenumber $k_1$ in the dashed blue curve on the left, there exists another wavenumber $k_2$ within the dashed blue region on the right that together to a fitting error which is smaller than the error of the Monte-Carlo simulation (the standard error of the mean). For more details see \Cref{app:MC methodology}. }
    \label{fig:2kps}
\end{figure}

\textbf{At least two effective wavenumbers.} The result of fitting for two effective wavenumbers for $ka = 0.36$ is shown in \Cref{fig:2kps}. The main conclusion is that this is the first clear evidence that there are two complex effective wavenumbers by using Monte-Carlo simulations which are highly accurate. In more detail: the figure shows that there are two separate effective wavenumbers: one within the blue dash curve on the left and the other, necessarily, within the blue dashed curve on the right. Wavenumbers within these dashed regions lead to fitting errors which are less than the error committed by the Monte-Carlo simulation. We also see the dispersion equation~\eqref{eqn:planewave-dispersion}, together with the Percus-Yevick pair correlation, predicts at least one wavenumber within the blue dash curve. The predicted wavenumbers when using the Hole-Correction pair-correlation is also shown. The dispersion equation \eqref{eqn:planewave-dispersion} predicts an infinite number of effective wavenumbers, so we present only the two wavenumbers with smallest imaginary part.

\begin{figure}[ht]
    \begin{subfigure}{0.49\textwidth}
        \centering
        \includegraphics[width=0.99\linewidth]
        {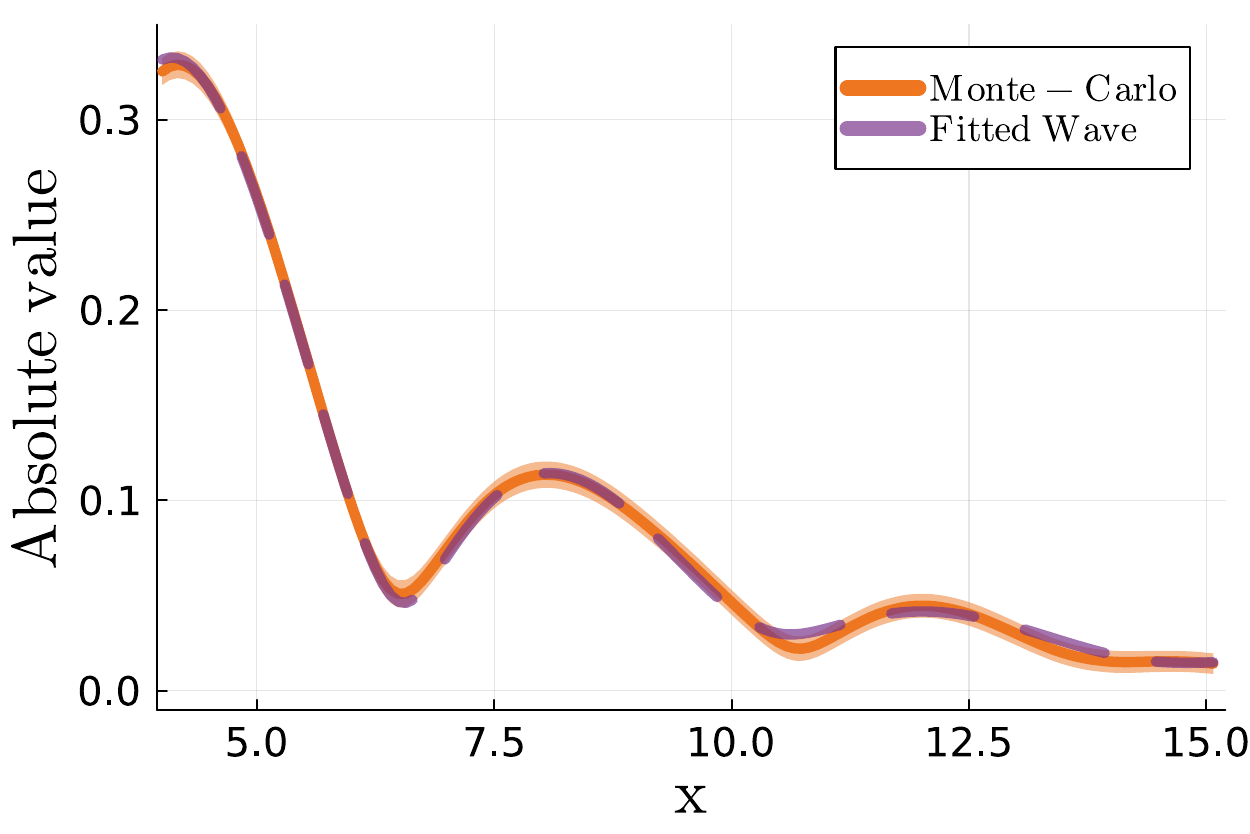}
        \caption{ }
        \label{fig:3kps_to_2kps_fit}
    \end{subfigure}
    \begin{subfigure}{.49\textwidth}
        \centering
        \includegraphics[width=0.99\linewidth]
        {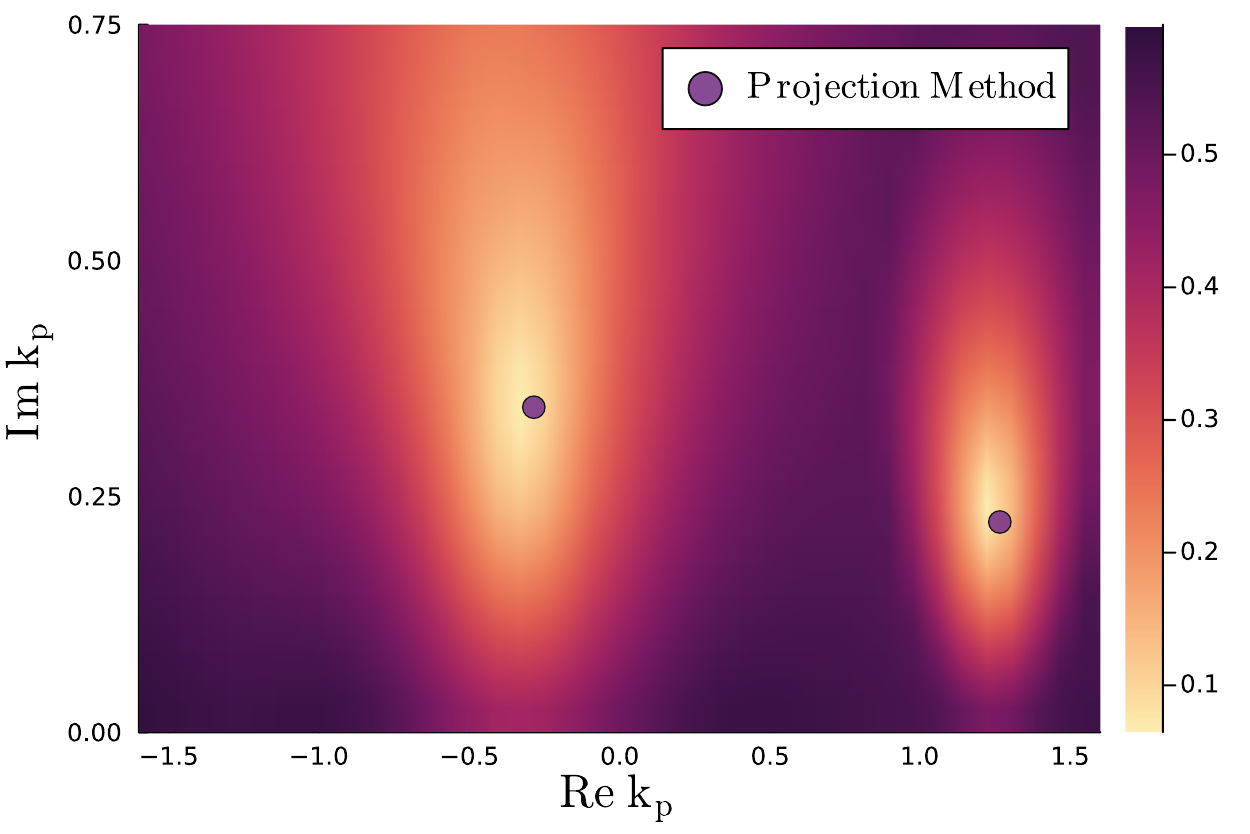}
        \caption{ }
        \label{fig:3kps_to_2kps_heatmap}
    \end{subfigure}
    \begin{subfigure}{.49\textwidth}
        \centering
        \includegraphics[width=0.99\linewidth]{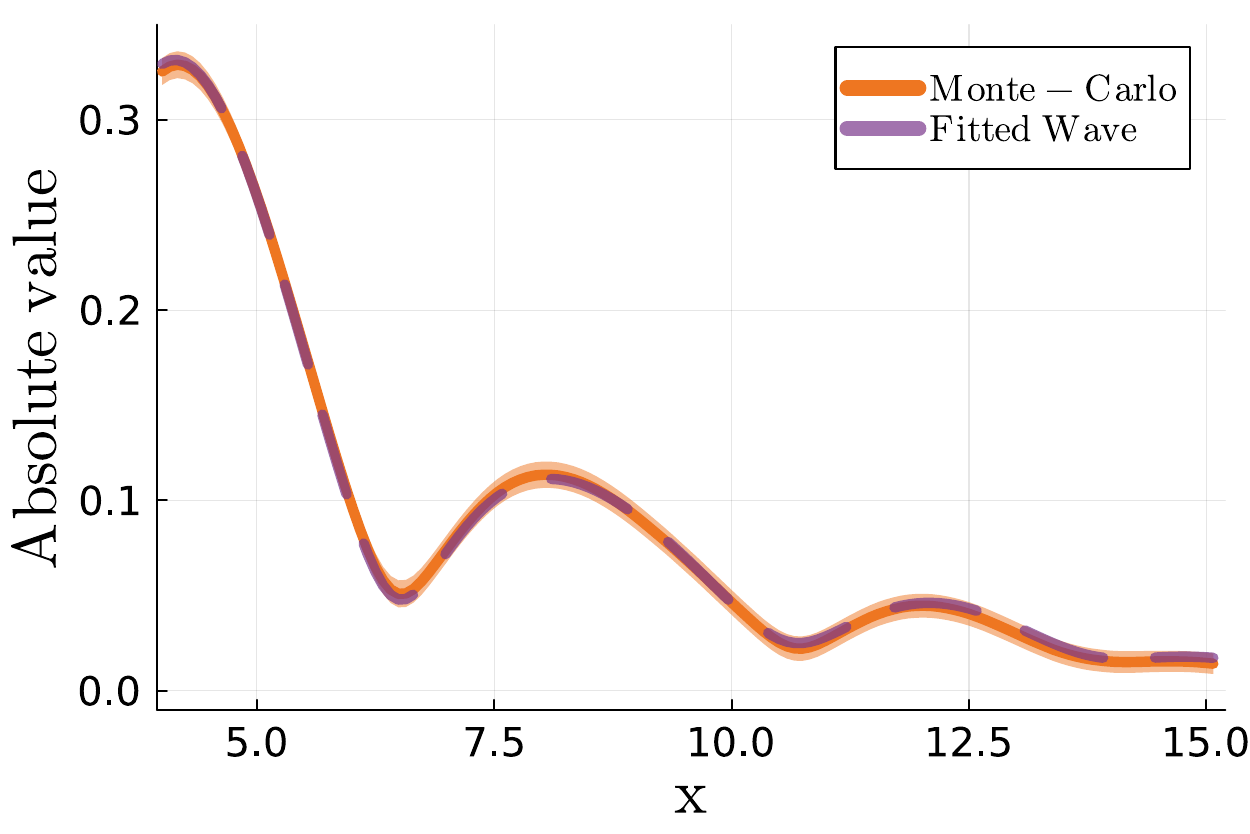}
        \caption{ }
        \label{fig:3kps_fit}
    \end{subfigure}
    \begin{subfigure}{.49\textwidth}
        \centering
        \includegraphics[width=0.99\linewidth]{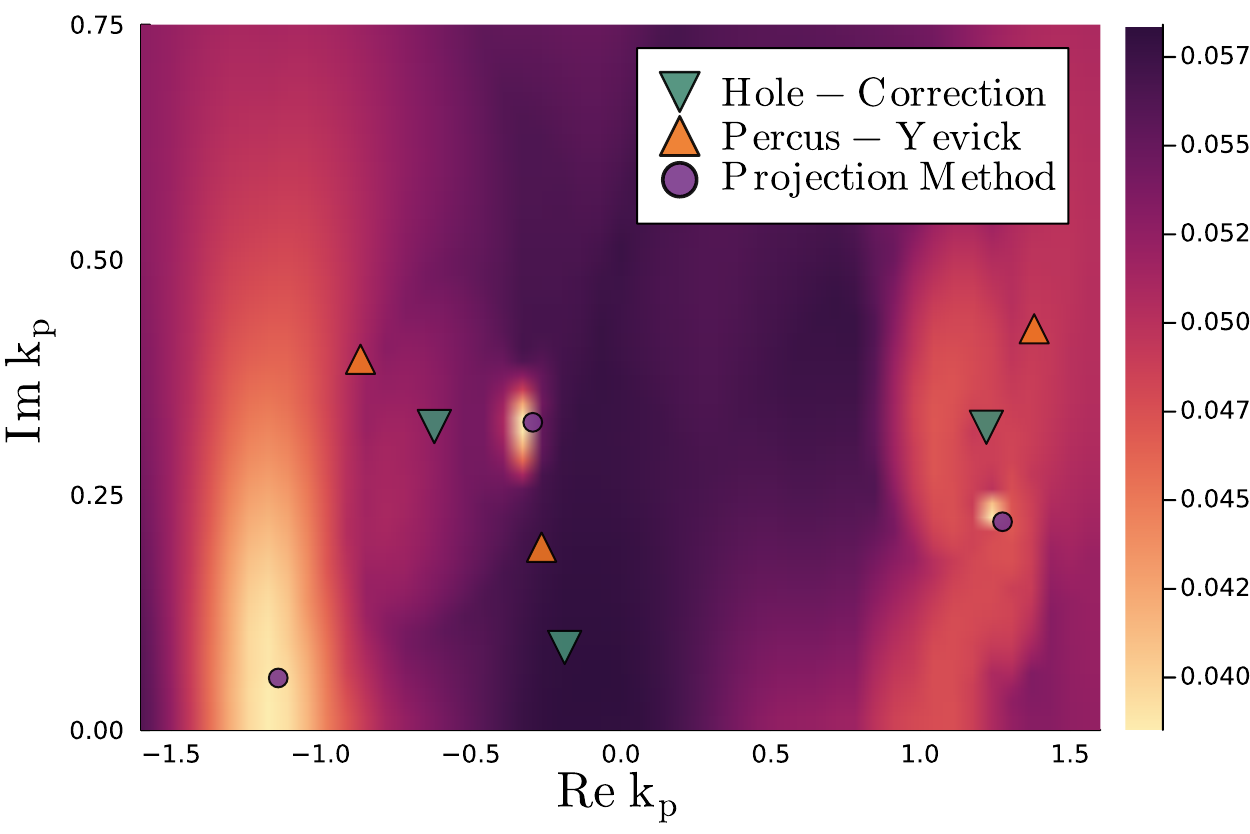}
        \caption{ }
        \label{fig:3kps_heatmap}
    \end{subfigure} 
  
        \caption{ The top two graphs show the result of using two effective wavenumbers in the formula \eqref{eq:effective_waves} to fit to the Monte-Carlo results for the particle properties $\rho_0 = c_0 = 0.30, \ \phi = 25 \%$ and frequency $k a = 0.62$. The graph \ref{fig:3kps_to_2kps_heatmap} shows a density plot over the effective wavenumbers, with light regions indicating that those wavenumbers better fit the data. However, all possible choices of two wavenumbers lead to fitting errors which are less than the standard error of the mean of the Monte-Carlo simulations. The two wavenumbers with the best fit are denoted by the \emph{Projection method}, and lead to the field shown in the graph \ref{fig:3kps_to_2kps_fit}. The bottom two graphs use three effective wavenumbers in the formula \eqref{eq:effective_waves} to fit to the Monte-Carlo results. In this case, we find 4 sets of wavenumbers, all close to each other, that have a fit error less than the standard error of the mean. The result of using the three wavenumbers with the best fit is shown in \ref{fig:3kps_fit}. However, in \ref{fig:3kps_heatmap} we see that there are many choices for the wavenumbers which lead to small fitting errors. In particular, the Projection method wavenumber with the smaller imaginary part is sensitive to small changes in the the Monte-Carlo results.
        }
\label{fig:3kps_comparison}  
\end{figure}

\textbf{Three effective wavenumbers.}  The \Cref{fig:3kps_comparison} shows in steps our attempt to identify three effective wavenumbers for $ka = 0.62$. The summary is that the amplitude $A_3$ for the third wavenumber in \eqref{eq:effective_waves} is too small, $|A_3| \approx 0.004$, and as a result, we can not reliable say whether the Monte-Carlo simulations show that there are indeed three effective wavenumbers. This is because the expected errors from the Monte-Carlo simulation are on the order of around $0.003$. However, we do conclude again that one effective wavenumber is not enough, at least two are needed.  

In more detail, the top two graphs of \Cref{fig:3kps_comparison} show the result of fitting the formula \eqref{eq:effective_waves} with $P=2$ to the Monte-Carlo results. Although there are clearly choices of two wavenumbers which fit the data well, there are no possible choices which lead to fitting errors that are less than the standard error of the mean, as shown by \Cref{fig:3kps_to_2kps_fit} which uses the two wavenumbers with the best fit. The bottom two graphs \Cref{fig:3kps_comparison} show how the fitting errors decrease when adding a third wavenumber, i.e. using $P=3$ in the formula \eqref{eq:effective_waves}. By using $P=3$, instead of $P=2$, the fitting error decreased from 6\% to 3.8\%. Although the fitting error does decrease to below the standard error of the mean, it is only a small decrease, and close to the errors inherent in the Monte-Carlo data. We note that the two Projection Method wavenumbers shown in \Cref{fig:3kps_to_2kps_heatmap} are equal to two of the Projection Method wavenumbers shown in \Cref{fig:3kps_to_2kps_heatmap}. 


\section{Deducing the average transmitted wave}
\label{sec:deducing transmission}

In \cite{gower2021effective,gower2019multiple,gower2019proof} the authors demonstrated that there exists several effective wavenumbers, however it was not clear how these appear in the average transmitted field. Taking inspiration from \cite{martin2011multiple}, the average transmitted field should be a sum of waves, each with a different effective wavenumber. In this section we demonstrate this for any incident field and material geometry.

We use the same notation given in~\cite{gower2021effective} and combine the methods shown in \cite{gower2021effective} and \cite{martin2011multiple} to show that the average transmitted field is a sum of the incident field plus several effective fields. 

We note that although the paper \cite{gower2021effective} is written for three dimensional particles, the results that lead up to \cite[Section 5]{gower2021effective} are valid for any dimension as long as we appropriately define the spherical waves $\mathrm u_n$ and $\mathrm v_n$ and the translation matrices $\mathcal V_{nn'}$ and $\mathcal U_{nn'}$. In the case of two dimensions and scalar waves these terms are
\begin{equation} \label{eqn:spherical waves 2D}
    \mathrm v_{n}(k\rv) =  \mathrm J_n(k r) \eu^{\iu m \theta},  \quad
    \mathrm u_{n}(k\rv) =  \mathrm H_n(k r) \eu^{\iu m \theta},
\end{equation}
\begin{equation} \label{eqn:translation matrices 2D}
    \mathcal V_{n n'}(\rv) =  \mathrm{v}_{n-n'}(\rv), \quad \text{and} \quad \mathcal U_{n n'}(\rv) =  \mathrm{u}_{n-n'}(\rv),
\end{equation}
where $\mathrm J_n$ and $\mathrm H_n$ are the Bessel function and Hankel function of the first kind, and $(r,\theta)$ are the polar coordinates of $\rv$. In our calculations here, and below, we will use the general notation rather than substitute the specific form for the two dimensions as shown above. This way, the proofs we present are valid for any dimension. 

\begin{figure}[H]

    \centering
    
    \begin{tikzpicture}
    \centering
        \draw (0,0) node {\includegraphics[width=0.7\textwidth ]{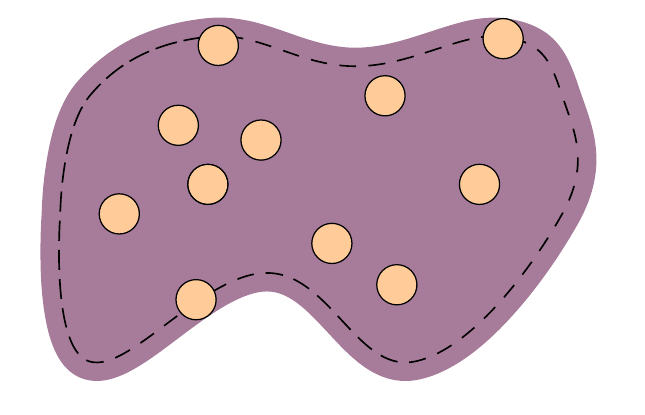} } ;
 	\draw(3.0, 2.77) node {$\mathcal{P}_{1}$};
 	\draw (-1.9,0.35) node {$\mathcal{P}_{2}$};
        \draw (-2.07, -1.55) node {$\mathcal{P}_{3}$};
        \draw (-2.37, 1.35) node {$\mathcal{P}_{4}$};
        \draw (1.05, 1.85) node {$\mathcal{P}_{5}$};
        \draw (-1.0, 1.1) node {$\mathcal{P}_{6}$};
        \draw (1.25, -1.3) node {$\mathcal{P}_{7}$};
        \draw (0.15, -0.6) node {$\mathcal{P}_{8}$};
        \draw (-1.73, 2.67) node {$\mathcal{P}_{9}$};
        \draw (2.6, 0.35) node {$\mathcal{P}_{10}$};
        \draw (-3.35, -0.1) node {$\mathcal{P}_{11}$};
        \draw (0.0, 0.25)  node {\Large $\mathcal{R} \setminus \mathcal{P}$};
        \draw (-5.7, 0.25) node {\Large $\mathcal{R}$};
        \draw (-3.2, -1.25) node {\Large $\mathcal{R}_{1}$};
        \draw [{Latex[length=2.5mm]}-](-4.3,-1.25) -- (-3.5,-1.25);
        \draw [-{Latex[length=2.5mm]}](-5.5,0.25) -- (-4.65,0.25);
    \end{tikzpicture}
    
    \caption{A two-dimensional region $\mathcal{R}$ filled with equal-sized disks $\mathcal{P}_j$, which represent the particles. The region $\mathcal{P} = \cup_j \mathcal{P}_j$ depicts the region inside the particles while the shaded region ($\mathcal{R} \setminus \mathcal{P}$) depicts the region outside the particles. The region $\reg$ completely contains all the particles, while the region $\reg_1$ contains only the particle centres. As the particles are at least one radius $a$ away from the boundary of $\reg$, we have that $\reg_1$ is smaller than $\reg$. Note that as the particles do not overlap, we have that $\mathcal{R} \setminus \mathcal{P} = \mathcal{R} \setminus \mathcal{P}_\ell - \sum_{j\not = \ell} \mathcal{P}_j$ for every $\ell$. }
    
    \label{fig:regions}  
\end{figure}
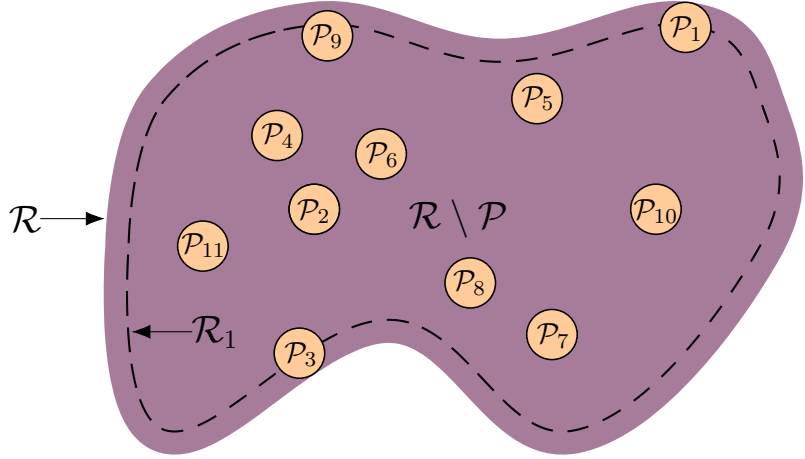

For just one configuration of particles, the way we represent the total field $u(\rv)$ depends on whether $\rv$ is inside a particle or not. We choose to write the field in the form ~\cite{Kristensson2015a,Kristensson2016,martin2011multiple}
\begin{equation} \label{eqn:total_transmitted}
    u(\rv) =
    \left\{\begin{alignedat}{2}
      &\ui(\rv) + \us(\rv), \quad && \text{for} \;\; \rv \in \reg \setminus \particle,
      \\
      & \uin^j(\rv), && \text{for} \;\; \rv \in \particle_j,
    \end{alignedat}
    \right.
\end{equation}
where $\ui(\rv)$ is the incident wave, $\us(\rv)$ is a sum of all the scattered waves, and $\uin^j(\rv)$ is the field inside particle $j$. We use $\particle_j$ to denote the region occupied by particle $j$, whereas $\particle = \cup_j \particle_j$ is the union of all particles. See \Cref{fig:regions} for an illustration.

The sum of the scattered waves $\us(\rv)$, shown in~\eqref{eqn:total_transmitted}, is given by
\begin{equation} \label{eqn:scattered wave}
  \us(\rv) =  \sum_{j=1}^J \us^j(\rv), \quad \text{where} \quad  \us^j(\rv) = \sum_n f_n^j \mathrm u_n (k \rv - k \rv_j).
\end{equation}
The field inside the $j$-th particle can be written in terms of a regular radial waves expansion:
\begin{equation} \label{eqn:internal field}
  \uin^j(\rv) = \sum_n b_n f_n^j \mathrm v_n(k_o \rv - k_o \rv_j),
\end{equation}
where 
\begin{equation} 
  b_n = \frac{T_n \H_n(k a) + \J_n(k a)}{T_n \J_n (k_o a)},
  \end{equation}
and, as a reminder, $T_n$ is given by \eqref{eqn:circular_t-matrix}. For a light introduction on the T-matrix and multiple scattering see \href{https://github.com/JuliaWaveScattering/MultipleScattering.jl/blob/master/docs/src/maths/multiplescattering.pdf}{this note} and \href{https://github.com/JuliaWaveScattering/MultipleScattering.jl/blob/master/docs/src/maths/acoustics.pdf}{this note}.

    
    
  

The ensemble average of any field $\mathrm f$, and the conditional ensemble average, is defined as 
\begin{align}\label{eqn:ensemble-average}
  &\ensem{\mathrm f}= \int_{\reg_1^{J}} \mathrm f \, p(\rv_1,\ldots, \rv_J) \mathrm d \rv_1 \cdots \mathrm d \rv_J,
  \\ \label{eqn:cond-ensemble-average}
  &\ensem{\mathrm f}(\rv_1) = \int_{\reg_1^{J-1}} \mathrm f \, p(\rv_2,\ldots, \rv_J| \rv_1) \mathrm d \rv_2 \cdots \mathrm d \rv_J,
\end{align}
where $\reg_1^{J}$ represents that the integration domain of all $J$ integrals is $\reg_1$, and $p(\rv_1,\rv_2,\ldots, \rv_J)$ is the probability density of having particles centered at $\rv_1,\rv_2, \cdots, \rv_J$, while $p(\rv_2,\ldots, \rv_J| \rv_1)$ is the conditional probability density and can be defined as
\begin{equation} \label{eqn:conditional average}
p(\rv_2,\ldots, \rv_J| \rv_1) := \frac{p(\rv_1,\ldots, \rv_J)}{p( \rv_1)}.    
\end{equation}

To calculate the ensemble average of the transmitted field, it is helpful to  write the field in the following form
\begin{equation} \label{eqn:total_transmitted2}
    u(\rv) =  [\ui(\rv) + \us(\rv)] \chi_{\reg \setminus \particle}(\rv)  + \sum_j \uin^j(\rv) \chi_{\particle_j}(\rv),
\end{equation}
where 
\begin{equation}
  \chi_{\mathcal A}(\rv) = \begin{cases}
  & 1 \quad \text{if} \;\; \rv \in \mathcal A,
  \\
  & 0 \quad \text{if} \;\; \rv \not \in \mathcal A.
\end{cases}
\end{equation}

By taking the ensemble average of \eqref{eqn:total_transmitted2} we obtain
\begin{equation} \label{eqn:ensem total field}
    \ensem{u(\rv)} =  \ensem{\ui(\rv) \chi_{\reg \setminus \particle}(\rv)}  + \ensem{\us(\rv) \chi_{\reg \setminus \particle}(\rv)}  + J \ensem{\uin^1(\rv) \chi_{\particle_1}(\rv)},
\end{equation}
where each term of the sum in~\eqref{eqn:total_transmitted2} is the same after ensemble averaging because they are all integrated over the same domain, which means that all particles are indistinguishable from each other. To calculate~\eqref{eqn:ensem total field} we need to make a few assumptions. 

\textbf{Isotropy and homogeneity assumption.} We assume isotropy and homogeneity, which means that $p(\rv_1)$, the probability density of one particle being at $\rv_1$, is a constant. Then, because the integral of $p(\rv_1)$ over $\reg_1$ has to equal 1, we conclude that
\begin{equation} \label{def:p r1}
    p(\rv_1) = \frac{1}{|\reg_1|} = \frac{\numdensity}{J},
\end{equation}
where $|\reg_1|$ is the volume of $\reg_1$, and $\numdensity$ is the number density of particles defined by
\begin{equation} \label{def:numdensity phi}
    \numdensity := \frac{J}{|\reg_{1}|} = \frac{\phi}{ |\particle_1|},
\end{equation}
with $\phi$ being the particle volume fraction, and $|\particle_1|$ being the volume of a particle.

Isotropy and homogeneity also imply that \cite{torquato2002random} the pair-correlation $g$ had the form~\eqref{def:pair-correlation-g}. For convenience, we now use the form
\begin{equation} \label{def:conditional_probability}
    p(\rv_2 | \rv_1) = \frac{\numdensity}{J - 1} g(|\rv_1 - \rv_2|).
\end{equation}
The term $(J - 1)$ appears now, rather than $J$, due to there being a finite number of particles, and it ensures that $g(|\rv_1 - \rv_2|) \to 1$ when particles become uncorrelated, as confirmed by \cite[Equation (8.1.2)]{kong2004scattering}. For more details on the pair-correlation see \cite{kong2004scattering}.    

\textbf{Correlation distance assumption.} We can only resolve the integrals appearing in \eqref{eqn:ensem total field} when $\rv$ satisfies
\begin{equation} \label{ass:r distance}
\min_{\rv_1 \in \partial\reg_1}\limits | \rv - \rv_1| \geq b_{12} + a,
\end{equation}
that is, when the distance of $\rv$ to the boundary $\partial \reg_1$ is greater than $b_{12} + a$. The distance $b_{12}$ appears in~\eqref{def:pair-correlation-g} and is sometimes called the correlation length. Our analysis shows that when the above is true, the transmitted field~\eqref{eqn:ensem total field} is a sum of effective waves, and the incident wave is no more.


\textbf{Quasi-crystalline assumption.} To calculate \eqref{eqn:ensem total field} we assume the closure assumption known as the Quasi-Crystalline Approximation (QCA). Details are given in~\cite{gower_reflection_2018,gower2021effective, Ma&Varadan1984}.

The calculations needed to simplify the ensemble averages in~\eqref{eqn:ensem total field} are shown in \Cref{sec:trans internal}-\ref{sec:trans scat}. To summarise, with all given assumptions, \eqref{eqn:ensem total field} leads to
\begin{multline} \label{eqn:ensem total field result 1}
    \ensem{u(\rv)} =  (1 - \phi) \ui(\rv) + J\ensem{ \us^1(\rv) \chi_{\reg \setminus \particle_1}(\rv) } -  J (J-1) \ensem{ \us^1(\rv) \chi_{\particle_2}(\rv) }
    \\
     + J \ensem{\uin^1(\rv) \chi_{\particle_1}(\rv)},
\end{multline}
where the terms above are given by~\eqref{eqn:internal field ensem}, \eqref{eqn:scat integral 0}, and \eqref{eqn:scat integral 3}. In the next section we show the above is a sum of waves.

\subsection{Transmitted effective waves} 
\label{sec:trans effective}

In this section, we use the effective wave assumption to demonstrate that~\eqref{eqn:ensem total field result 1} is composed of functions which satisfy effective wave equations. Below we show that
\begin{equation} \label{eqn:transmit split}
  \ensem{u(\rv)} = w_\text{inc}(\rv) + \sum_{p = 1}^ \infty w_p(\rv),
\end{equation}
where 
\begin{equation} \label{eqns:helmholtz}
\nabla^2  w_\text{inc}(\rv) + k^2  w_\text{inc}(\rv) = 0 \quad \text{and} \quad \nabla^2 w_p(\rv) + k_p^2 w_p(\rv) = 0.    
\end{equation}
We also show that $w_\text{inc}(\rv) = 0$ when \eqref{ass:r distance} holds, and when using the effective boundary condition which is deduced from first principals in~\cite{gower2021effective}. That is, the incident wave is not present inside the material\footnote{We note that $k$ can not equal $k_p$ \cite{gower_reflection_2018}. }. 


For what follows, to keep the notation concise, we define the ball region using standard set-builder notation
\begin{equation} \label{def:ball}
  \mathcal B(\vec x; R) = \left \{\vec y \in \mathbb R^2  : \;|\vec x - \vec y| \leq R \right \}.
\end{equation}

The first term in~\eqref{eqn:ensem total field result 1} is given by $(1 - \phi) \ui(\rv)$, which clearly contributes to the term $w_\text{inc}(\rv)$ in \eqref{eqn:transmit split}, because the incident wave satisfies the Helmholtz equation \eqref{eqns:helmholtz}${}_1$.

The second term $J\ensem{ \us^1(\rv) \chi_{\reg \setminus \particle_1}(\rv) }$ is more involved. In \Cref{sec:trans scat}  we prove this term has the reduced form \eqref{eqn:scat integral 0}. To calculate this term, first we use~\eqref{eqn:scattered wave} together with~\eqref{eqn:cond-ensemble-average} to obtain
\begin{equation} \label{eqn:us 1 conditional}
  \ensem{\us^1(\rv)}(\rv_1) = \sum_n \ensem{f_n}(\rv_1) \mathrm u_n (k \rv - k \rv_1).
\end{equation}

Following the method shown in \cite{gower2021effective}, we use a series expansion of the form 
\begin{equation} \label{eqn:effective-assumption}
    \ensem{f_n}(\rv_1) = \sum_p f_{p,n}(\rv_1), 
\end{equation}
where each $f_{p,n}$ satisfies a Helmholtz equation: $\nabla^2 f_{p,n}(\rv_1) + k_p^2 f_{p,n}(\rv_1) = 0$,  and the $k_p$ and $f_{p,n}$ are determined from the governing equation of $\ensem{f_n}(\rv_1)$. In \cite{gower2021effective} it is shown how such a series solution can be used for any material region, where \cite{gower2019proof} proofs the above form for plane-waves. 

Using the effective wave series expansion~\eqref{eqn:effective-assumption} in \eqref{eqn:scat integral 0} leads to
\begin{equation}
J\ensem{ \us^1(\rv) \chi_{\reg \setminus \particle_1}(\rv) } =   \numdensity \sum_{np}\int_{\reg_1 \setminus \mathcal B(\rv;a)}  f_{p,n}(\rv_1) \mathrm u_n (k \rv - k \rv_1) \mathrm d \rv_1.
\end{equation}
As shown in \cite[Section 4]{gower2021effective}, we can use Green's second identity to write
\begin{equation} \label{eqn:greens identity}
  \int_{\reg_1 \setminus \mathcal B(\rv;a)}  f_{p,n}(\rv_1) \mathrm u_n (k \rv - k \rv_1) \mathrm d \rv_1 =  \frac{\mathcal I_{p,n}(\rv) - \mathcal J_{p,n}(\rv)}{k^2 - k_p^2},
\end{equation}
where
\begin{align}
  & \mathcal I_{p,n}(\rv) = \int_{\partial \reg_1}  \frac{\partial  f_{p,n}(\rv_1)}{\partial \vec \nu_1} \mathrm u_n (k \rv - k \rv_1) -  f_{p,n}(\rv_1) \frac{\partial\mathrm u_n (k \rv - k \rv_1)}{\partial \vec \nu_1} \mathrm d A_1,
  \\
  & \mathcal J_{p,n}(\rv) = \int_{\partial \mathcal B(\vec 0, a)}  \frac{\partial  f_{p,n}(\rv  - \vec x_1)}{\partial \vec \nu_1} \mathrm u_n (k \vec x_1) -  f_{p,n}(\rv  - \vec x_1) \frac{\partial\mathrm u_n (k \vec x_1)}{\partial \vec \nu_1} \mathrm d A_1,
\end{align}
from which we can see that
\[
\nabla^2 \mathcal I_{p,n}(\rv) + k^2 \mathcal I_{p,n}(\rv) = 0 \;\; \text{and} \;\; \nabla^2 \mathcal J_{p,n}(\rv) + k_p^2 \mathcal J_{p,n}(\rv) = 0.
\]
So, clearly, $\mathcal I_{p,n}(\rv)$ contributes to $w_\text{inc}(\rv)$, while $\mathcal J_{p,n}(\rv)$ contributes to $w_p(\rv)$ in~\eqref{eqn:transmit split}.

The third term in~\eqref{eqn:ensem total field result 1} is given by~\eqref{eqn:scat integral g 2} in \Cref{app:isotropic pair}, which after using \eqref{eqn:effective-assumption} and \eqref{eqn:us 1 conditional} becomes
\begin{multline}
J(J-1)\ensem{ \us^1(\rv) \chi_{\particle_2}(\rv) } =   \phi \numdensity \sum_{np}\int_{\reg_1 \setminus \mathcal B(\rv; b_{12} -a)}  f_{np}(\rv_1) \mathrm u_n (k \rv - k \rv_1) \mathrm d \rv_1
\\
 - \numdensity^2  \sum_{np} \int_{\mathcal B(\vec 0;b_{12}+a) \setminus \mathcal B(\vec 0;a_{12}-a)}   f_{p,n}(\rv - \vec x_1) \mathrm u_n (k \vec x_1)  G(\vec x_1)  \mathrm d \vec x_1,
\end{multline}
where $G(\vec x_1)$ is defined just below~\eqref{eqn:scat integral g 2}, although it is not required for our goals here. 


The first of the above integrals is analogous to~\eqref{eqn:greens identity}, so leads to terms of the form~\eqref{eqn:transmit split}. The second of these integrals only has $\rv$ dependence in $f_{p,n}(\rv - \vec x_1)$, and therefore contributes to $w_p(\rv)$ in \eqref{eqn:transmit split}.

The fourth term in~\eqref{eqn:ensem total field result 1} is given by~\eqref{eqn:internal field ensem}, which after using~\eqref{eqn:internal field}, \eqref{eqn:effective-assumption} and the change of variables from $\rv_1$ to $\vec x_1 = \rv - \rv_1$ becomes
\begin{equation}
\label{eq:intermediate_computation}
J \ensem{\uin^1(\rv) \chi_{\particle_1}(\rv)}  = \numdensity  \sum_{pn} b_n \int_{\mathcal B(\vec 0;a)} f_{p,n}(\rv - \vec x_1) \mathrm v_n(k_o \vec x_1)\mathrm d \rv_1,
\end{equation}
which can only contribute to terms of the form $w_p(\rv)$ in \eqref{eqn:transmit split}.

\subsection{The average of the incident field}

In the previous section we demonstrated that~\eqref{eqn:ensem total field} is a sum of terms which satisfy the background and effective wave equations as shown in \eqref{eqn:transmit split}. The term $w_\text{inc}$, that satisfies the background wave equation, can be seen as what remains of the incident field. In much of the literature \cite{Carminati2021, Fearn1996, kong2004scattering, Lax1952, Tishkovets2011} it is simply assumed that $w_\text{inc} := 0$. This is often called  the ``extinction theorem'', despite it being an assumption. In papers such as \cite{martin2011multiple} that calculate the average field from first principals, it is not clear that $w_\text{inc} = 0$. Here we remove any doubt by proving that when sufficiently inside the material, given by condition~\eqref{ass:r distance}, we have that $w_\text{inc} := 0$ for any incident field, any material region $\reg$, any frequency, and all types of particles. 

Using the results from \Cref{sec:trans effective}, we collect the terms in \eqref{eqn:ensem total field result 1} that satisfy the background wave equation to obtain
\begin{equation} \label{eqn:average incident}
  w_\text{inc}(\rv) = (1 - \phi) \ui(\rv) + \sum_{pn} \frac{\numdensity(1 - \phi)}{k^2 - k_p^2} \mathcal I_{p,n}(\rv)
\end{equation}

Below we show that the right side of the above is zero by using the ensemble boundary condition given by \cite[Equation (4.8)]{gower2021effective} 
\begin{equation} \label{eqn:boundary condition}
\sum_{n'}
 \mathcal{V}_{n'n}(k\rv_1)g_{n'}   +
  \numdensity \sum_{nn'p}\frac{\mathcal I_{p,n'n} (\rv_1)}{k^2 - k^2_p}  = 0,
\end{equation}

where
\begin{equation} \label{eqn:boundary integral}
\mathcal I_{p,n'n} (\rv_1) = \int_{\partial \reg_1 }  \mathcal{U}_{n'n}(k\rv_1 - k\rv_2) \frac{\partial f_{p,n'}(\rv_2)}{\partial \vec\nu_2}
  - \frac{\partial \mathcal{U}_{n'n}(k\rv_1 - k\rv_2) }{\partial \vec\nu_2} f_{p,n'}(\rv_2) \mathrm dA_2,
\end{equation}
with $\rv_2$ being the variable of integration and $\vec \nu_2$ the normal to the boundary $\partial \reg_1$, and the $g_{n}$ are the coefficients of the incident wave
\begin{equation} \label{eq:incident wave expansion}
    \ui(\rv) = \sum_{n} g_n \mathrm v_n(k \rv). 
\end{equation}
The above assumes the source of the incident wave is outside of the region where the particles are \cite{gower2021effective}. The other assumptions needed to deduce the ensemble boundary conditions \eqref{eqn:boundary condition} are the same assumptions we have used for the calculations in this paper, except the boundary condition is only valid when 
\begin{equation} \label{cond:boundary conditoin}
\min_{\rv_2 \in \partial \reg_1} |\rv_1  - \rv_2| \geq a_{12}.    
\end{equation}

To start the demonstration, we multiply both sides  of \eqref{eqn:boundary condition}  by $\mathrm v_n(k \rv - k \rv_1)$ and sum over $n$ to obtain
\begin{equation} \label{eqn:ensemble boundary}
\sum_{nn'}
 \mathcal{V}_{n'n}(k\rv_1)g_{n'} \mathrm v_n(k \rv - k \rv_1)  +
  \numdensity \sum_{nn'p}\frac{\mathcal I_{p,n'n} (\rv_1)}{k^2 - k^2_p}  \mathrm v_n(k \rv - k \rv_1) = 0.
\end{equation}
The above can be simplified by using the fundamental property of translation matrices that
\begin{equation}\label{eq:translation_spherical_waves}
 \left\{\begin{aligned}
&\mathrm{v}_n(k\rv + k \dv) = \sum_{n'}\mathcal{V}_{nn'}(k\dv)\mathrm{v}_{n'}(k\rv),\quad \text{ for all }\dv
   \\
&\mathrm{u}_n(k\rv + k\dv) = \sum_{n'} \mathcal{U}_{nn'}(k\dv)\mathrm{v}_{n'}(k\rv),\quad |\rv|<|\dv|
   \\
 \end{aligned}\right..
 \end{equation}

Using the property of the translation matrices above and \eqref{eq:incident wave expansion},  we see that 
\begin{equation} \label{eqn:incident translation}
  \sum_{n n'}
   \mathcal{V}_{n'n}(k\rv_1)g_{n'} \mathrm v_n(k \rv - k \rv_1) = \sum_n g_n \mathrm{v}_n (k \rv) = 
   \ui(\rv),
\end{equation}
in accordance with \cite[Equation (2.3)]{gower2021effective}.

Next, by choosing $\rv$ such that \eqref{ass:r distance} is satisfied, it is then possible to choose $\rv_1$ so that the condition~\eqref{cond:boundary conditoin} is true and such that 
\[
|\rv - \rv_1| < |\rv_1 - \rv_2| \quad \text{for every} \quad \rv_2 \in \partial \reg_1
\]
This enables us to use the translation property \eqref{eq:translation_spherical_waves} in~\eqref{eqn:boundary integral} to obtain
\begin{multline}\label{eqn:boundary integral translation}
\sum_n \mathcal I_{p,n'n} (\rv_1) \mathrm v_n(k \rv - k \rv_1) =
\\
\int_{\partial \reg_1 }  \sum_n \mathrm v_n(k \rv - k \rv_1) \mathcal{U}_{n'n}(k\rv_1 - k\rv_2) \frac{\partial f_{p,n'}(\rv_2)}{\partial \vec\nu_2}
  - \sum_n \mathrm v_n(k \rv - k \rv_1)  \frac{\partial \mathcal{U}_{n'n}(k\rv_1 - k\rv_2) }{\partial \vec\nu_2} f_{p,n'}(\rv_2) \mathrm dA_2
  \\
  = \int_{\partial \reg_1 }  \mathrm u_{n'}(k\rv - k\rv_2) \frac{\partial f_{p,n'}(\rv_2)}{\partial \vec\nu_2}
    - \frac{\partial \mathrm u_{n'}(k\rv - k\rv_2) }{\partial \vec\nu_2} f_{p,n'}(\rv_2) \mathrm dA_2 =
    \mathcal I_{p,n'} (\rv).
\end{multline}

Substituting~\eqref{eqn:incident translation} and \eqref{eqn:boundary integral translation} into \eqref{eqn:ensemble boundary} leads to
\begin{equation} \label{eqn:ensemble boundary v2}
\ui(\rv) + \numdensity \sum_{n'p}\frac{\mathcal I_{p,n'} (\rv)}{k^2 - k^2_p} = 0.
\end{equation}
Finally, substituting~\eqref{eqn:ensemble boundary v2} into~\eqref{eqn:average incident}, we conclude the extinction theorem $w_\text{inc}(\rv) = 0$ for $\rv$ that satisfies~\eqref{ass:r distance}. That is, there is no term in the average transmitted wave that satisfies the background wave equation.

\section{Conclusions}
\label{sec: conclusions}

The initial goal of this work was to find clear evidence that there exists at least two effective wavenumbers in an averaged particulate material. It is highly unusual to have two different wavenumbers for an isotropic homogeneous media supporting only scalar waves. However theoretical works \cite{gower2019proof,gower2019multiple,gower2021effective,JRWillis2020,willis2022some,willis2023transmission} have predicted the existence of at least two effective wavenumbers, and their presence changes the overall transmitted and reflected waves.

\textbf{Monte Carlo results.} To verify the existence of multiple effective wavenumbers we used very precise simulations that calculated scattered waves from different particle configurations and then took an average over the different particle configurations. This turned out to be far more computational expensive then we expected, and required extensive and careful analysis. To summarise, \Cref{fig:2kps} clearly shows that there are two separate wavenumbers that contribute to the field, and that the wavenumbers predicted by the Monte-Carlo method are similar to the wavenumbers predicted by the theory.   

\textbf{When it matters.} A natural question that appeared during this work was how to find the parameters that led to multiple effective wavenumbers. That is, for what scenarios will the classical theory that uses only one effective wavenumber \cite{Carminati2021,linton_multiple_2005,linton_multiple_2006,martin2011multiple,mishchenko2014electromagnetic,Varadan1983} be inaccurate? Previous work \cite{gower2019multiple,gower2021effective} demonstrated that there is a dispersion equation~\eqref{eqn:planewave-dispersion} which provides the effective wavenumbers $k_p$, and that if there is only one wavenumber $k_1$ with an imaginary part much small than all others Im $k_1 \ll $ Im $k_p$ for $p=2,\ldots$, then the classical theory will be accurate. 

However, solving the dispersion equation~\eqref{eqn:planewave-dispersion} can be time consuming, specially for higher frequencies and wide range of parameters. When plotting the regions where multiple wavenumbers appear we saw a clear pattern shown in \Cref{fig:soft-phase-diagram}: particles which are strong scatterers lead to multiple effective wavenumbers. In \Cref{fig:soft-phase-diagram} the green curves show the scattering strength of just one particle \eqref{eqn:scattering strength}. Finding the parameters that lead one particle to be a strong scatterer is far more practical than solving the dispersion equation~\eqref{eqn:planewave-dispersion}, and proved to be a surprisingly good measure. To summarise: strong multiple scattering triggers multiple effective wavenumbers. 

\textbf{Resonators.} In the field of metamaterials, strong scatterers such as resonators are often used to tailor the overall behaviour of the material \cite{craster2012acoustic}. Using this strategy for disordered, or random, particulates will lead to multiple effective wavenumbers, and will complicate how to predict the overall properties of the material. To truly understand the effect of these resonators it is first necessary to plot their dispersion diagrams by solving the dispersion equation \eqref{eqn:planewave-dispersion} with the T-matrix $T_n$ depending on the type of particle used. An example of such a diagram is given by \Cref{fig:attenuations}. 


\textbf{The theoretical results.} When deciding how best to sample the transmitted field, we realised that within the theoretical formulation for ensemble averaging particulates, it was not clear that the transmitted wave is a sum of waves with effective wavenumbers. This led us to derive the missing results and provide a general proof about the incident and transmitted waves.   

\textbf{Proof of extinction.} It is often assumed that the average field inside a random material does not contain any remnant of the incident wave. This is called the Ewald-Oseen extinction theorem, but as far as the authors are aware, there is no proof of this conjecture for particulate materials. In this work, we were able to prove this extinction theorem for any particulate (for scalar isotropic waves), any frequency, and material geometry. The proof is given in \Cref{sec:deducing transmission}, with the final equation that proves extinction being \eqref{eqn:ensemble boundary v2}. The proof also provides the extinction length: the distance required for the incident wave to travel with the material until it is extinct. We proved that the extinction length is equal to the correlation length plus the particle radius $b_{12} + a$, see equation~\eqref{def:pair-correlation-g} where these quantities are relative to the pair-correlation.        

\textbf{Proof of transmitted effective waves.} The same proof for extinction also served to prove that the average transmitted field is a sum of effective waves, when the distance from the material boundary is greater than the extinction length. The proof is shown in \Cref{sec:deducing transmission}. We note, that particularly in the field of continuously varying random media \cite{Carminati2021,vynck2021light}, it is assumed that the average transmitted field satisfies an effective wave equation. By proving this for particulates, from a microscopic approach, we provide a link between the two approaches. 

\textbf{Future avenues.} This work shows that using Monte-Carlo simulations to approximate a semi-infinite media, such as a plate, filled with particles, is still computational challenging. We feel that future work focused on validating effective theories for particulates should focus on finite materials (in the computational sense), such as a cylinder filled with cylindrical particles and a sphere filled with spherical particles. There is a theoretical framework to validate against \cite{gower2021effective}. In terms of theoretical developments, our work has shown a connection between the particulate microscopic approach to effective waves \cite{gower2021effective,mishchenko2014electromagnetic,mishchenko_multiple_2006,Tishkovets2011,Tsang2000} and approaches for continuously varying random media \cite{Carminati2021,vynck2021light}. That is, we demonstrate the effective wave series used in the macroscopic approach given by \eqref{eqn:effective-assumption} does lead to the average transmitted wave being a sum of effective waves, as illustrated by \eqref{eqn:transmit split}. We believe the calculations we provide now paves the way to answer the follow open question: are the two approaches equivalent?  



\section*{Author contributions}
AK conceived of study, drafted the manuscript, wrote all the code for the numerical calculations, developed the theoretical calculations, and produced all the figures. ALG helped conceive the study, edited the manuscript, assisted with and verified the theoretical calculations. PSP assisted with the projection method and with the figures related with this method, verified the calculations, and edited the manuscript.

\section*{Data and reproducability}
\label{sec: data}

To produce our results we used the open source software \cite{2020MultipleScatering.jl}, \cite{2018EffectiveWaves.jl}, \cite{Optim.jl-2018}.

\section*{Acknowledgements}

The authors would like to acknowledge Kevish Napal for insightful and helpful discussions. Aristeidis Karnezis gratefully acknowledges travel support from the UK Acoustics Network (EP/V007866/1, EP/R005001/1). Artur Gower gratefully acknowledges support from EPSRC (EP/V012436/1). Paulo Piva gratefully acknowledges funding from a Case studentship with Johnson Matthey.  

\appendix

\section{The Monte-Carlo methodology}
\label{app:MC methodology}

Here we present more details on how we performed the Monte-Carlo simulations, and analysed the results. For a reference on Monte-Carlo methods we refer to the book \cite{Tsang2001}. 

To compute the ensemble average wave \eqref{def:ensemble average} with Monte-Carlo simulations, the waves scattered by particles within a plate geometry (as shown in \Cref{fig:plate_geometry}) have to be simulated tens of thousands of times, with each simulation having hundreds of particles, before the standard error of the mean converges \cite{martin_multiple_2006}. For each simulation we calculate exactly how the incident wave $\ui(x) = \eu^{\iu k x}$ scatters from all the particles. For these reasons, careful considerations are needed to determine how to perform the Monte-Carlo simulations.

In order, we explain how we created each particle configuration, how we determined the plate width and height, including consideration of convergence, and finally, how we analyzed the data.

\textbf{Sequential addition.} To place the particles we use the strategy of Sequential addition as described in Chapter 8, Section 2 of \cite{Tsang2001}. In essence, we place one particle at a time according to a random uniform distribution. If the particle overlaps with another particle it is rejected. The process is repeated until we obtain a desired particle volume fraction $\phi$.

\textbf{The plate width.} 
Choosing an appropriate width $W$ for the plate $\mathcal R$ was based on two factors: 
\begin{itemize}
    \item The result from the theory shown by \eqref{eq:effective_waves}  predicts the plate needs to have width $W > 2a + 2b_{12}$ for $\ensem{u(x)}$ to be exactly equal to a sum of effective waves. The minimum value for $b_{12}$ is $a_{12}$, which we found to be accurate enough for our tolerances. Using $b_{12} = a_{12} > 2a$  implies that we need a plate with $ W > 6a = 7.2$, as we used $a=1.2$ for all numerical experiments.
    \item The plate width $W$ can not be too wide, otherwise the average wave $\ensem{u(x)}$ will be completely attenuated, which is a computational waste. Also, in the region where the wave is completely attenuated it is impossible to estimate the $k_p$ by fitting the formula \eqref{eq:effective_waves}. Materials, and frequencies, that lead to \eqref{eq:effective_waves} needing more than one effective wavenumber $k_1$ to accurately approximate $\ensem{u(x)}$ are highly attenuating materials. See Figure \ref{fig:average_comparison} for an illustration of the region where we fit the formula \eqref{eq:effective_waves}. 
\end{itemize}

\textbf{The plate height.} If the plate filled with particles, as shown in \Cref{fig:plate_geometry}, was infinite in height, and the particles were excited by a plane-wave, then the average wave $\ensem{u(x)}$ would be exactly a sum of plane-waves given by \eqref{eq:effective_waves}. See \cite{gower2021effective} for details. In practice, it is not of course possible to exactly simulate the wave scattered from an infinite plate filled with one specific arrangement of particles. The approximation often used is to have a cell filled with a random set of particles, and then to use periodic tilling of this cell \cite{chekroun_comparison_2009}. To avoid the artifacts produced by periodic tilling we perform a convergence study to determine at what height a plate filled with particles behaves approximately like a plate of infinite size.  

Let $H$ be the height of the plate, which is illustrated in \Cref{fig:plate_geometry}. To determine the size needed for $H$, we first choose a large value $H = 600$, with the plate width $W = 20$, and then filled this plate with one configuration of particles $\Lambda$, according to the sequential addition method. We then calculate the total wave at
\begin{equation}
    \label{eq:measuring_region}
    u \left(x, 0 \right) \quad \text{for the values} \quad 2a \leq x \leq W - 2a,
\end{equation}
to create a function $\mathbf U_{600}(x)$. To determine the influence of $H$, we then reduce the height. For example, we use $H = 590$ and remove from $\Lambda$ any particles that are now outside of the box with the reduced height. We then take the updated configuration of particles $\Lambda$ and recalculate the total wave in the same region to create the function $\mathbf U_{590} (x)$.

For a range of heights we compute the relative error
\begin{equation}
    \label{eq:error_plate}
    \text{Error \%} = 100 \frac{\| \mathbf U_{H}(x) - \mathbf U_{600}(x) \|}{ \| \mathbf U_{600}(x) \|}.
\end{equation}
By calculating theses errors for a range of heights we can plot the error against the height as shown in \Cref{fig: convergence} below. 


For sound-soft particles, see \Cref{table:particle properties} for details, and frequency $ka = 0.3$, we see that the errors have converged. This means that $\mathbf U_{600}(x)$ is approximately the same as $\mathbf U_{\infty}(x)$, and we can estimate that the height of the plate has to be approximately $H = 400$ for the scattered waves to have an error of less than 1\% in comparison to $\mathbf U_{\infty}(x)$.

For sound-hard particles, see \Cref{table:particle properties} for details, and $ka = 0.3$, the errors have not converged as can be seen from \Cref{fig: convergence}, even for very larger heights. This means it is unclear what is the error relative to an infinite plate $\mathbf U_\infty$. For this reason, we do not focus on this case, and only performed one simulation with $H=400$. Our hypothesis is that the plane-wave when scattered from the corner of the plate leads to a transmitted wave that travels down inside the plate relatively unobstructed, as these types of particles are weak scatterers.   

When fitting for the effective wavenumbers, we need to consider how the errors due to truncating the height of the plate may affect the fitting and the acceptable fit errors.

\begin{figure}[H]

\centering

\begin{subfigure}{0.49\textwidth}
    \includegraphics[width=\textwidth]{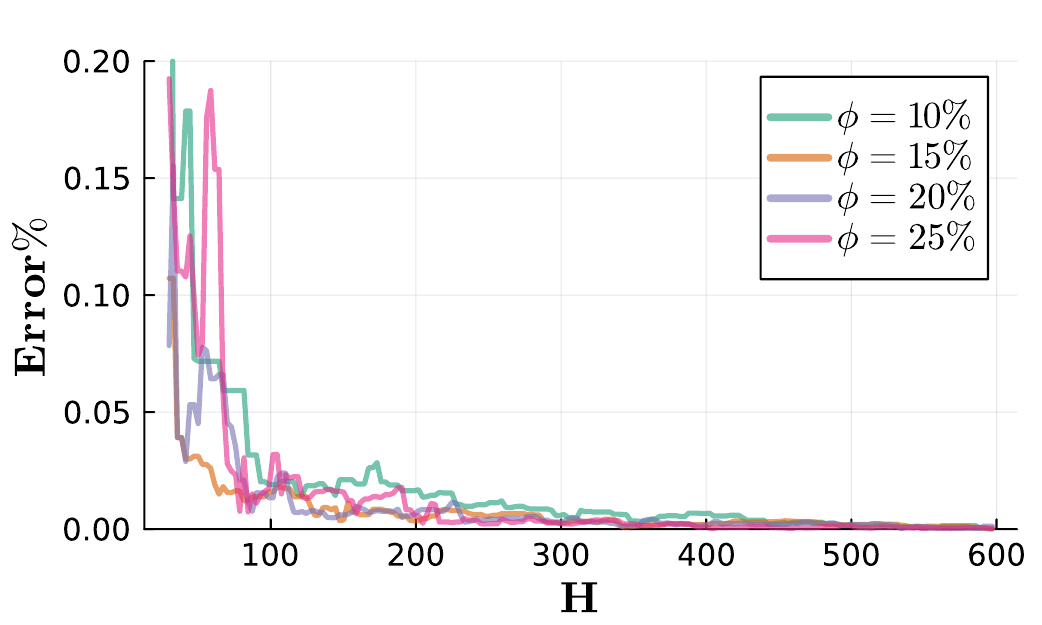}
    \caption{Sound-soft scatterers}
    \label{fig:sound_soft_convergence}
\end{subfigure}
\hfill
\begin{subfigure}{0.49\textwidth}
    \includegraphics[width=\textwidth]{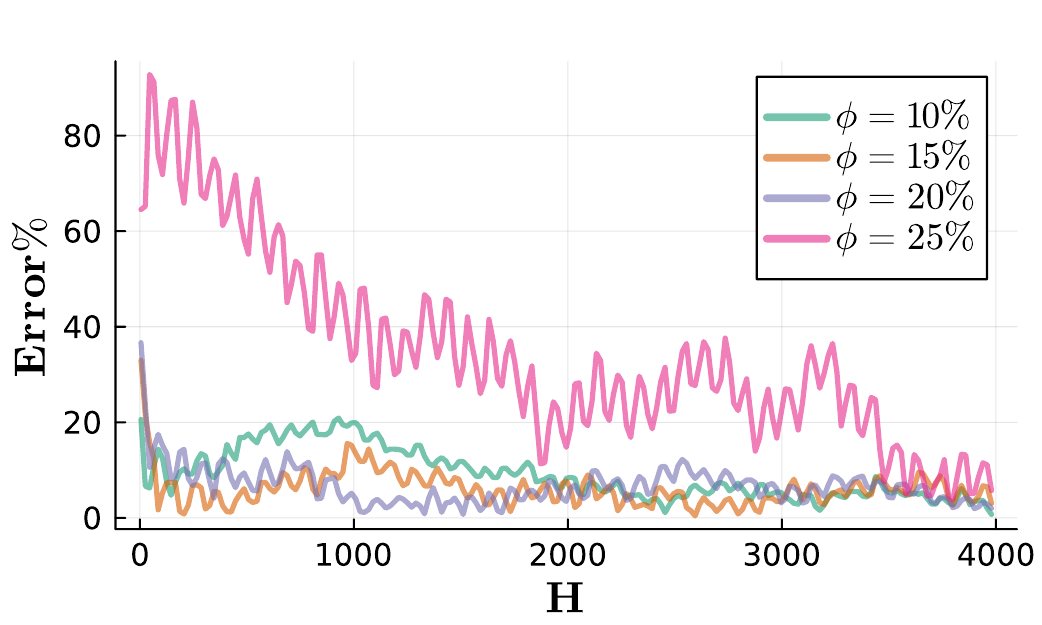}
    \caption{Sound-hard scatterers}
    \label{fig:sound_hard_convergence}
\end{subfigure}
        
\caption{Shows the rate of convergence which can be described with the normalised difference between the scattered waves \eqref{eq:error_plate} with respect to the height of the plate $\textbf{H}$.}

\label{fig: convergence}
\end{figure}

\textbf{Analysing the data.} 
We generate Monte-Carlo simulations for the different cases presented in \Cref{subsec: MonteCarlo} with a plate of width $W = 20$ and height $H = 400$.
The data from Monte-Carlo results for the transmitted wave \eqref{eq:effective_waves} is of the form 
\begin{equation} 
\notag
\braket{u(x,0 ; \Lambda)} = \sum_{p=1}^P \left( A_p^{+} \E^{\I k_p x} 
 + A_p ^{-} \E^{-\I k_p x} \right) 
+ \epsilon (x), \quad \text{for} \;\;  2a \leq x \leq W-2a,
\end{equation}
where $\epsilon (x)$ is a small error that falls inside the standard error of the mean. For the cases where only one effective wavenumber was predicted, the average wave $\ensem{u(x,0;\Lambda)}$ is fitted well by \eqref{eqn:average_wave_1}, using non-linear optimization libraries in Julia \cite{Optim.jl-2018}.

For every wavenumber $k_p$ there are potentially two waves: one travelling towards the positive $x$ - direction and another travelling in the negative $x$ - direction. For the cases where the average wave is completely attenuated when it reaches the edge of the plate, $x = 20$, we should have $A_p^- = 0$, suppressing the wave travelling on the negative x-direction.

From the Monte-Carlo data for cases with more than one effective wavenumber, we find that $\ensem{u(x,0;\Lambda)}$ is below the standard error of the mean for $x \geq 15$. For this reason, we can take $A_p^- = 0$ and fit for only for $A_p^+$ and $k_p$. That is we fit to the Monte-Carlo data functions of the form:
\begin{equation}
    \label{eq:fitting_curve}
    h(x) = \sum_{p=1}^P A_p^+ \E^{\I k_p x}, \quad \text{for} \;\;  4 \leq x \leq 15,
\end{equation}
with $\mathrm{Im}[k_p] > 0$. In the case presented in \Cref{fig:average_wave_2kps} the transmitted wave attenuates quicker, therefore we focus the fitting on a narrower range, $4\leq x \leq 10$.

Another reason to use \eqref{eq:fitting_curve}, is that fitting for $A_p^+$, $A_p^-$, and $k_p$ in the case of multiple effective wavenumbers can become ill-posed. This is related to how inverting a Laplace transform is ill-posed, see Section 2.1 of \cite{kaipio2006statistical}. That is, the more terms included in the sum shown in \eqref{eq:effective_waves} the more ill-posed is the problem of recovering the $k_p$, $A_p^+$, and $A_p^-$ from data of the average wave $\ensem{u(x, 0; \Lambda)}$.

For the cases with more than one effective wavenumber, non-linear optimization \cite{Optim.jl-2018} is unstable. Therefore, we implement a method, the \emph{Projection method}, to both fit the formula \eqref{eq:fitting_curve} and estimate the sensitivity of the parameters $k_p$ to the data.


\textbf{Projection Method.}
To fit Monte-Carlo data for the cases with more than one effective wavenumbers, we design an algorithm that sweeps over all possible values of the wavenumbers $k_p$ and for each case performs a linear fit, by using least-squares, to predict the amplitudes $A_p^+$ in \eqref{eq:fitting_curve}. For that, we consider a mesh in the complex plane $\mathcal C \subset \mathbb C$ from which we sample values of $k_p$. A sketch of the algorithm for the case $P = 3$ is given below in Algorithm \ref{alg:Proj_Meth}.   

\begin{algorithm}[hbt!]
\caption{Algorithm for the Projection Method}\label{alg:Proj_Meth}
\KwData{$\ensem{u(x,0;\Lambda)}$}
\For{all values of $k_1 \in \mathcal C$}{
    \For{all values of $k_2, k_3 \in \mathcal C$}{
        \If{$k_1 \neq k_2 \neq k_3$}{
            Determine the values of $A_p^+$ that best fit $\ensem{u(x,0;\Lambda)}$ using least-squares \;
            Using \eqref{eq:fitting_curve}, compute the error $\varepsilon(k_1, k_2, k_3) = \| h(x) - \ensem{u(x,0;\Lambda)} \|$ \;
            Store $\varepsilon(k_1, k_2, k_3)$;
        }
    }
Store $\varepsilon(k_1) = \min_{k_2,k_3} \varepsilon(k_1, k_2, k_3)$
}
\KwResult{$\varepsilon(k_1)$}
\end{algorithm}

The algorithm calculates the minimum error $\varepsilon(k_1)$ for every possible $k_1$, and it is this error which is shown in \Cref{fig:2kps,fig:3kps_to_2kps_heatmap,fig:3kps_heatmap}. We perform a further analysis to establish if the optimal wavenumbers are located at the local minima shown, and whether the fitted curves when using \eqref{eq:fitting_curve} are within the standard error of the mean of the Monte-Carlo simulation. 
The blue region in \Cref{fig:2kps}, contains all wavenumbers for which the fitted curves are within the standard error of the mean. There are clearly two distinct disconnected regions for possible values of $k_p$, which shows that the transmitted wave is composed of a sum of two effective waves.

\paragraph{Computational time.} Our study utilizes high-fidelity Monte-Carlo simulations, whose computational cost was significant. This limits the number of cases we could investigate with Monte-Carlo. For example, we needed to execute 40,000 simulations for scenarios depicted in Figures \ref{fig:average_comparison}, \ref{fig:2kps} and \ref{fig:3kps_comparison}. Each simulation involve hundreds of particles, and are computationally demanding especially at higher frequencies denoted by $ka$. To manage this, we employed parallel processing across multiple processors. For instance, simulating sound-soft particles at a frequency of $ka$ = 0.36 requires about 72 hours for completion on an 11th Gen. Intel Core i7 with 8 cores. So without parallel processing, the runtime would increase by at least eight-fold. 

\section{The ensemble average transmission}

Below we reduce the terms in~\eqref{eqn:ensem total field} to reach equation~\eqref{eqn:ensem total field result 1}. For this section we use the notation and assumptions introduced in \Cref{sec:deducing transmission}. 

In many calculations throughout this section, for any function $\mathrm f$ which depends on particle configuration, we use that
\begin{multline} \label{eqn:ensemble to r1 integral}
    J \ensem{\mathrm f} = J \int_{\reg_1^J} \mathrm f p(\rv_1, \ldots, \rv_J) \mathrm d \rv_1 \cdots \mathrm d \rv_J 
    \\
    = \numdensity \int_{\reg_1^J}\mathrm f p(\rv_2, \ldots, \rv_J| \rv_1) \mathrm d \rv_1 \cdots \mathrm d \rv_J 
    = \numdensity  \int_{\reg_1} \ensem{\mathrm f}(\rv_1) \mathrm d \rv_1,
\end{multline}
where we used, in order, the definitions \eqref{eqn:ensemble-average} and \eqref{eqn:conditional average} and \eqref{def:p r1}. If the $\mathrm f$ only depends on $\rv_1$, then we further have that $ \ensem{\mathrm f}(\rv_1) = \mathrm f$ because
\[
 \int_{\reg_1^{J-1}} p(\rv_2,\ldots, \rv_J | \rv_1)  \mathrm d \rv_2  \cdots \mathrm d \rv_J = 1,
\]
which is true for any joint probability density.

\subsection{The transmitted internal field} \label{sec:trans internal}

We start by calculating the simplest term. Using \eqref{eqn:internal field}, the definitions of the ensemble average \eqref{eqn:cond-ensemble-average} and \eqref{eqn:ensemble-average}, then \eqref{eqn:ensemble to r1 integral} leads to
\begin{equation} \label{eqn:internal field ensem}
J \ensem{\uin^1(\rv) \chi_{\particle_1}(\rv)} 
=  \numdensity \int_{\mathcal B(\rv;a)} \ensem{\uin^1(\rv)}(\rv_1)\mathrm d \rv_1,
\end{equation}
where the ball $\mathcal B(\rv;a)$ is defined by \eqref{def:ball}, and we use the assumption~\eqref{ass:r distance}.

\subsection{The transmitted incident field} \label{sec:trans incident}
The next simplest computation is the ensemble average of the incident wave term in~\eqref{eqn:total_transmitted2}. To do this we will demonstrate the following equalities:
\begin{equation}
  \ensem{\ui(\rv) \chi_{\reg \setminus \particle}(\rv)} = \ui(\rv) \ensem{\chi_{\reg \setminus \particle}(\rv)} =  \ui(\rv) \ensem{1 - \chi_{\particle}(\rv)} = \ui(\rv) (1 - \phi),
\end{equation}
where $\phi$ is the particle volume fraction defined by~\eqref{def:numdensity phi}.

First, we use the ensemble average~\eqref{eqn:ensemble-average}, then take $\ui(\rv)$ outside of the integrals, as it does not depend on the particle positions, to reach
\begin{equation}
   \ensem{\ui(\rv) \chi_{\reg \setminus \particle}(\rv)} 
  \\
  = \ui(\rv) \ensem{\chi_{\reg \setminus \particle}(\rv)}.
\end{equation}
To calculate the ensemble average on the right we use
\begin{equation} \label{eqn:chi negative}
\chi_{\reg \setminus \particle}(\rv) = 1 - \chi_{\particle}(\rv),
\end{equation}
leading to
\begin{equation}
  \ensem{\chi_{\reg \setminus \particle}(\rv)} = 
  \ensem{1}  - \ensem{\chi_{\particle}(\rv)} 
  = 1 -  \sum_j \ensem{\chi_{\particle_j}(\rv)} 
\end{equation}
where we used the definition that integrating a probability density function $p$ over all its variables gives 1, and
\begin{equation} \label{eqn:additive chi}
  \chi_{\particle}(\rv) p(\rv_1,\rv_2,\ldots, \rv_J) = \sum_{j=1}^J \chi_{\particle_j}(\rv)p(\rv_1,\rv_2,\ldots, \rv_J),
\end{equation}
which holds because if any two particles overlap, we have that $p(\rv_1,\rv_2,\ldots, \rv_J) =0$. Therefore, if $\chi_{\particle_j}(\rv) = 1$, then $\chi_{\particle_\ell}(\rv) = 0$ for $\ell \not = j$.

Next, we use that particles are indistinguishable, except for their positions, so after ensemble averaging $\ensem{\chi_{\particle_j}(\rv)} = \ensem{\chi_{\particle_1}(\rv)}$ for every $j$, which together with \eqref{eqn:ensemble to r1 integral} leads to
\begin{equation}
\ensem{\chi_\particle(\rv)} = \sum_{j=1}^J\ensem{\chi_{\particle_j}(\rv)} =  J\ensem{\chi_{\particle_1}(\rv)} 
= \numdensity \int_{\reg_1} \chi_{\particle_1}(\rv) \mathrm d \rv_1 
= \numdensity \int_{\mathcal B(\rv,a)} \mathrm d \rv_1 = \phi,
\end{equation}
where we used~\eqref{def:ball} and \eqref{ass:r distance}.

\subsection{The transmitted scattered field} \label{sec:trans scat}
The most involved term to calculate in \eqref{eqn:ensem total field} is $\ensem{\us(\rv) \chi_{\reg \setminus \particle}(\rv)}$, which will require that we demonstrate the following steps
\begin{multline} \label{eqn:ensem scat overview}
  \ensem{\us(\rv) \chi_{\reg \setminus \particle}(\rv)} = \sum_j \ensem{ \us^j(\rv)  \chi_{\reg \setminus \particle}(\rv)} =   J\ensem{ \us^1(\rv)  \chi_{\reg \setminus \particle}(\rv)} =   J\ensem{ \us^1(\rv)  \prod_{j=1}^J \chi_{\reg \setminus \particle_j}(\rv) }
  \\
  =  J\ensem{ \us^1(\rv) \chi_{\reg \setminus \particle_1}(\rv) } -  J\sum_{j=2}^J \ensem{ \us^1(\rv) \chi_{\particle_j}(\rv) }
  \\
  =  J\ensem{ \us^1(\rv) \chi_{\reg \setminus \particle_1}(\rv) } -  J (J-1) \ensem{ \us^1(\rv) \chi_{\particle_2}(\rv) }.
\end{multline}

The first three equalities above are a result of using, in order, \eqref{eqn:scattered wave}, that particles are indistinguishable, and
\[
\chi_{\reg \setminus \particle}(\rv) p(\rv_1,\rv_2,\ldots, \rv_J) = \prod_{j=1}^J \chi_{\reg \setminus \particle_j}(\rv) p(\rv_1,\rv_2,\ldots, \rv_J),
\]
%
which is a result of $p(\rv_1,\rv_2,\ldots, \rv_J) = 0$ when any two particles overlap. The non-overlapping of particles also leads to
\[
\chi_{\reg \setminus \particle_1}(\rv) \prod_{j=2}^J \chi_{\reg \setminus \particle_j}(\rv) p(\rv_1,\rv_2,\ldots, \rv_J)  = \left[ \chi_{\reg \setminus \particle_1}(\rv)  - \sum_{j=2}^J \chi_{\particle_j}(\rv) \right] p(\rv_1,\rv_2,\ldots, \rv_J),
\]
which we use to conclude the second line in \eqref{eqn:ensem scat overview}, and the third line is just a result of particles being indistinguishable again.

Below we further simplify the last two terms in~\eqref{eqn:ensem scat overview}. Using \eqref{eqn:ensemble to r1 integral} we can reach
\begin{equation}\label{eqn:scat integral 0}
  J\ensem{ \us^1(\rv)  \chi_{\reg \setminus \particle_1}(\rv) } =
  \numdensity \int_{\reg_1}  \ensem{\us^1(\rv)}(\rv_1)\chi_{\reg \setminus \particle_1}(\rv)  \mathrm d \rv_1 = \numdensity \int_{\reg_1 \setminus \mathcal B(\rv;a)}  \ensem{\us^1(\rv)}(\rv_1) \mathrm d \rv_1 ,
\end{equation}
where $\ensem{\us^1(\rv)}(\rv_1)$ is given by~\eqref{eqn:us 1 conditional}.

For the last term in~\eqref{eqn:ensem scat overview}, we further use
\[
 p(\rv_2,\ldots, \rv_J | \rv_1) =  p(\rv_2| \rv_1) p(\rv_3,\ldots, \rv_J | \rv_1, \rv_2),
\]
\eqref{eqn:scattered wave}, \eqref{eqn:cond-ensemble-average}, and the definition of $\ensem{f_n}(\rv_1,\rv_2)$ given by  \cite[Equation (3.10) and (3.11)]{gower2021effective}, followed by 
analogous steps shown in \eqref{eqn:us 1 conditional}, to obtain
\begin{equation}\label{eqn:scat integral 1}
J \ensem{ \us^1(\rv) \chi_{\particle_2}(\rv) } =
  \numdensity\sum_n\int_{\reg_1^2}  \ensem{f_n}(\rv_1,\rv_2) \mathrm u_n (k \rv - k \rv_1)  p(\rv_2| \rv_1)  \chi_{\particle_2}(\rv) \mathrm d \rv_1  \mathrm d \rv_2.
\end{equation}
To simplify the above we first use the Quasi-Crystalline Approximation (QCA) \eqref{eqn:QCA}
\begin{equation} \label{eqn:QCA}
  \ensem{f_n}(\rv_1,\rv_2) \approx \ensem{f_n}(\rv_1),
\end{equation}
which is needed to deduce effective wavenumbers \cite{gower2021effective, linton_multiple_2005, Ma&Varadan1984}.

Before we show how to simplify \eqref{eqn:scat integral 1} for a general pair-correlation $g$, which we shown in \Cref{app:isotropic pair}, we first deduce the results for the simplest pair correlation called Hole-Correction. It is far easier to understand this case first. The Hole-Correction approximation is the result of taking $b_{12} = a_{12}$ in the general pair-correlation~\eqref{def:pair-correlation-g}.


Using \eqref{eqn:QCA} and~\eqref{def:conditional_probability}  in the integral~\eqref{eqn:scat integral 1}, and swapping the order of integration leads to
\begin{equation}\label{eqn:scat integral 2}
J (J-1) \ensem{ \us^1(\rv) \chi_{\particle_2}(\rv) }
= \numdensity^2\int_{\reg_1}   \ensem{\us^1(\rv)}(\rv_1) \int_{\reg_1} \chi_{\particle_2}(\rv) g(|\rv_1 - \rv_2|)
\mathrm d \rv_2 \mathrm d \rv_1.
\end{equation}
Next we use $b_{12} = a_{12}$ and~\eqref{def:pair-correlation-g} which implies that 
\[
g(|\rv_1 - \rv_2|) = \chi_{\reg_1 \setminus \mathcal B(\rv_1 ; a_{12})}(\rv_2), 
\]
which we substitute into \eqref{eqn:scat integral 2} to reach
\begin{equation}\notag
J (J-1) \ensem{ \us^1(\rv) \chi_{\particle_2}(\rv) }
= \numdensity^2\int_{\reg_1}   \ensem{\us^1(\rv)}(\rv_1) \int_{
\mathcal B(\rv;a) \setminus \mathcal B(\rv_1; a_{12})
}
\mathrm d \rv_2 \mathrm d \rv_1,
\end{equation}
where we used that $\mathcal B(\rv;a)$ is completely contained in $\reg_1$ for every $\rv_2$ due to~\eqref{ass:r distance}. There are values for $\rv_1$ for which the integral above is zero. To see this we note
\[
\rv_2 \in \mathcal B(\rv;a) \implies |\rv_2 - \rv| \leq a \quad \text{and} \quad  
\rv_2 \not \in \mathcal B(\rv_1; a_{12}) \implies|\rv_2 - \rv_1| > a_{12},
\]
then from the triangular inequality we have 
\[
|\rv_1 - \rv| \geq |\rv_1 - \rv_2| - |\rv_2 - \rv| >  a_{12} - a.
\]
The above implies that the region of integration for $\rv_1$ is just $\reg_1 \setminus \mathcal B(\rv; a_{12} - a)$. We now further split this region of integration into two disjoint regions: the first is $\reg_1 \setminus \mathcal B(\rv;a_{12}+a)$ and the second is $\mathcal B(\rv;a_{12}+a) \setminus \mathcal B(\rv;a_{12}-a)$.

For the first region  $\rv_1 \in \reg_1 \setminus \mathcal B(\rv;a_{12}+a)$ implies that  $|\rv_1 - \rv| \geq a_{12} + a $, which together with $\rv_2 \in \mathcal B(\rv;a)$ 
leads to
\[
|\rv_1 - \rv_2| \geq |\rv_1 - \rv| -  |\rv - \rv_2| > a_{12},
\]
due to the triangle inequality.
In other words, the region of integration for $\rv_2$ becomes $\rv_2 \in \mathcal B(\rv;a) \setminus \mathcal B(\rv_1; a_{12}) = \mathcal B(\rv;a)$.

For the second region $\rv_1 \in \mathcal B(\rv;a_{12}+a) \setminus \mathcal B(\rv;a_{12}-a)$, which implies that 
\[
a < a_{12} - a \leq |\rv_1 - \rv| \leq a_{12} + a.
\]
The above guarantees that the two spheres $\mathcal B(\vec r;a)$ and $\mathcal B(\rv_1; a_{12})$ will intersect. Let $\mathcal V$ be this region of intersection, then the region of integration of $\rv_2$ becomes $\mathcal B(\rv;a) \setminus \mathcal B(\rv_1; a_{12}) = \mathcal B(\rv;a) \setminus \mathcal V$. This is useful as $\mathcal V$ is formed of two spherical caps whose volume is easy to calculate\footnote{See the website \url{https://mathworld.wolfram.com/Sphere-SphereIntersection.html} for details.}.

Using the split of these two regions for $\rv_1$ we obtain
\begin{multline} \label{eqn:scat integral 3}
J(J-1) \ensem{ \us^1(\rv) \chi_{\particle_2}(\rv) } =
 \numdensity^2 \int_{\reg_1 \setminus \mathcal B(\rv;a_{12}+a)}  \ensem{\us^1(\rv)}(\rv_1) \int_{\mathcal B(\rv;a)} \mathrm d \rv_2 \mathrm d \rv_1
\\
+ \numdensity^2\int_{\mathcal B(\rv;a_{12}+a) \setminus \mathcal B(\rv;a_{12}-a)}   \ensem{\us^1(\rv)}(\rv_1)  \left[\frac{4}{3} \pi a^3 - V_\text{cap}(|\rv-\rv_1|) \right] \mathrm d \rv_1
\\
= \phi \numdensity \int_{\reg_1 \setminus \mathcal B(\rv;a_{12}-a)} \ensem{\us^1(\rv)}(\rv_1) \mathrm d \rv_1
\\
-  \numdensity^2\int_{\mathcal B(\rv;a_{12}+a) \setminus \mathcal B(\rv;a_{12}-a)}  \ensem{\us^1(\rv)}(\rv_1)  V_\text{cap}(|\rv-\rv_1|) \mathrm d \rv_1
\end{multline}
where $V_\text{cap}(d)$ is the volume of $\mathcal V$ and $d = |\rv - \rv_1|$. By using the formulas for spherical caps\footnote{See the website \url{https://mathworld.wolfram.com/Sphere-SphereIntersection.html} for details.} we can calculate that
\[
V_\text{cap}(d) = \frac{\pi}{ 12 d}(a + a_{12} - d)^2 (d^2 + 2 (a + a_{12}) d -3 (a - a_{12})^2 ).
\]

%

\subsubsection{An isotropic pair correlation}\label{app:isotropic pair}
In the previous section we choose a simple pair-correlation to simplify the integral \eqref{eqn:scat integral 2}. Here we show how to reduce this integral when assuming a more general form for the isotropic pair correlation give by \eqref{def:pair-correlation-g}.   

In the pair-correlation \eqref{def:pair-correlation-g} we assume there is a value $b_{12}$ for which $g(r) = 1$ when $r \geq b_{12}$. This is an approximation, but it is essential for the results in this section. this distance $b_{12}$, which is also called the correlation length, dictates at what distance inside the material the incident wave will be extinct.

Following closely the steps that led to~\eqref{eqn:scat integral 3}, we now split the integral over $\rv_1$ into two regions $\reg_1 \setminus \mathcal B(\rv;b_{12}+a)$ and $\mathcal B(\rv;b_{12}+a) \setminus \mathcal B(\rv; a_{12} - a)$, which leads to
\begin{multline} \label{eqn:scat integral g}
J(J-1) \ensem{ \us^1(\rv) \chi_{\particle_2}(\rv) } =
 \numdensity^2 \int_{\reg_1 \setminus \mathcal B(\rv;b_{12}+a)}  \ensem{\us^1(\rv)}(\rv_1) \int_{\mathcal B(\rv;a)} \mathrm d \rv_2 \mathrm d \rv_1
\\
+ \numdensity^2\int_{\mathcal B(\rv;b_{12}+a) \setminus \mathcal B(\rv;a_{12}-a)}   \ensem{\us^1(\rv)}(\rv_1)  \int_{\mathcal B(\rv; a)} g(|\rv_1 - \rv_2|)\mathrm d \rv_2 \mathrm d \rv_1,
\end{multline}
where we use \eqref{ass:r distance} to guarantee that the ball $\mathcal B(\rv; b_{12} + a)$ is completely contained within the region $\reg_1$. Without the condition \eqref{ass:r distance} it does not seem possible to show that the incident wave becomes extinct, so we hypothesise that this is a necessary condition, as well as sufficient. 

The integral on the right of the first line of~\eqref{eqn:scat integral g} was already resolved in the previous section, except now we replace $a_{12}$ with $b_{12}$. For the integrals on the second line of~\eqref{eqn:scat integral g}, we use the change of variables from $\rv_2$ to $\rv_{21} = \rv_2 - \rv_1$ and $\rv_1$ to $\vec x_1 = \rv - \rv_1$ to obtain
\begin{multline} \label{eqn:scat integral g 2}
J(J-1) \ensem{ \us^1(\rv) \chi_{\particle_2}(\rv) } =
 \numdensity \phi \int_{\reg_1 \setminus \mathcal B(\rv;b_{12}+a)}  \ensem{\us^1(\rv)}(\rv_1) \mathrm d \rv_1
\\
+ \numdensity^2\int_{\mathcal B(\vec 0;b_{12}+a) \setminus \mathcal B(\vec 0;a_{12}-a)}   \ensem{\us^1(\rv)}(\rv - \vec x_1) G(\vec x_1) \mathrm d \vec x_1,
\end{multline}
where
\[
G(\vec x_1) = \int_{\mathcal B(\vec x_1; a)} g(r_{21})\mathrm d \rv_{21}.
\]
and $r_{21} = |\rv_{21}|$. This concludes the calculations in this section.






\bibliography{references} 

\end{document}